\documentclass[pra,reprint,superscriptaddress]{revtex4-1}

\pdfoutput=1

\usepackage{amsmath,amssymb, amsfonts, graphicx, hyperref, natbib}
\usepackage[mathscr]{euscript}

\newcommand{\hH}{\hat{H}}
\newcommand{\ket}[1]{\left| #1 \right\rangle}
\newcommand{\bra}[1]{\left\langle #1 \right|}
\newcommand{\expect}[1]{\left\langle #1 \right\rangle}
\newcommand{\smallexpect}[1]{\langle #1 \rangle}
\newcommand{\hS}{\hat{S}}
\newcommand{\hB}{\hat{B}}

\begin{document}
\author{Bhuvanesh Sundar}
\thanks{BS (\href{mailto:bs55@rice.edu}{bs55@rice.edu}) and KCW (\href{mailto:kwang411@stanford.edu}{kwang411@stanford.edu}) contributed equally to this work}
\affiliation{Department of Physics and Astronomy, Rice University, Houston, Texas 77005, USA}
\affiliation{Rice Center for Quantum Materials, Rice University, Houston, Texas 77005, USA}

\author{Kenneth C. Wang}
\thanks{BS (\href{mailto:bs55@rice.edu}{bs55@rice.edu}) and KCW (\href{mailto:kwang411@stanford.edu}{kwang411@stanford.edu}) contributed equally to this work}
\affiliation{Department of Physics and Astronomy, Rice University, Houston, Texas 77005, USA}
\affiliation{Department of Physics, Stanford University, Stanford, California 94305, USA}

\author{Kaden R. A. Hazzard}
\email{kaden@rice.edu}
\affiliation{Department of Physics and Astronomy, Rice University, Houston, Texas 77005, USA}
\affiliation{Rice Center for Quantum Materials, Rice University, Houston, Texas 77005, USA}

\title{Analysis of continuous and discrete Wigner approximations for spin dynamics}
\date{\today}

\begin{abstract}
We compare the continuous and discrete truncated Wigner approximations of various spin models' dynamics to exact analytical and numerical solutions. We account for all components of spin-spin correlations on equal footing, facilitated by a recently introduced geometric correlation matrix visualization technique [R. Mukherjee {\em et al.}, Phys. Rev. A {\bf 97}, 043606 (2018)]. We find that at modestly short times, the dominant error in both approximations is to substantially suppress spin correlations along one direction.
\end{abstract}

\maketitle
\section{Introduction}
The dynamics of quantum matter is linked to several important phenomena in physics, such as thermalization or lack thereof~\cite{nandkishore2015many}, dynamical phase transitions~\cite{eckstein2009thermalization, heyl2018dynamical}, and universality in out-of-equilibrium dynamics~\cite{prufer2018observation, langen:ultracold_2015, polkovnikov:nonequilibrium_2011, lamacraft-moore:potential-insights_2012, altman2015non}. Understanding these phenomena is challenging, partly due to the lack of theoretical tools to accurately simulate them. There is an urgent need for such tools because recent experiments have made strides in measuring out-of-equilibrium dynamics; see, for example, Refs.~\cite{kim2010quantum, islam2013emergence, britton2012engineered, bohnet2016quantum, de2013nonequilibrium, de2016probing, zeiher2016many, mukherjee2016accessing, low2009universal, takei2016direct, guardado2018probing, nichols2019spin, lienhard2018observing, hazzard2014many, garttner2017measuring}. Several numerical methods, such as exact diagonalization~\cite{manmana2005time, rigol2008thermalization, prelovvsek2013ground, sandvik2010computational}, time-dependent density-matrix renormalization group~\cite{white2004real, daley2004time, vidal2004efficient, wolf2014solving, schuch2011classifying}, perturbative and Keldysh techniques~\cite{bray2002theory, calabrese2005ageing, henkel2008non, henkel2011non, kamenev2011field, tauber2014critical}, kinetic theories and phase-space methods~\cite{walls2007quantum, blakie2008dynamics, gardiner2004quantum, orioli2017nonequilibrium}, and numerical linked-cluster expansions~\cite{tang2013short, rigol2014quantum, white2017quantum, guardado2018probing, nichols2019spin, mallayya2018quantum}, have been used to calculate such dynamics. However, all these methods have limitations, ranging from being restricted to small or low-dimensional systems, to being accurate only for weakly interacting, close-to-equilibrium, or short-time situations.

In this paper we compare two popular and related semiclassical approximations for the dynamics of quantum matter, namely, the continuous truncated Wigner approximation (TWA) and discrete truncated Wigner approximations (DTWA)~\cite{wootters1987wigner, polkovnikov2003quantum, polkovnikov2010phase, schachenmayer2015many}, with each other and with exact analytical or numerical solutions. These approximations have been used frequently in recent years to simulate the dynamics of spin models~\cite{schachenmayer2015many, schachenmayer2015dynamics, pucci2016simulation, orioli2017nonequilibrium, orioli2018relaxation, babadi2015far, valtierra2017twa, czischek2018quenches, wurtz2018cluster}, which are some of the most ubiquitous dynamics probed in experiments~\cite{kim2010quantum, islam2013emergence, britton2012engineered, bohnet2016quantum, de2013nonequilibrium, de2016probing, zeiher2016many, mukherjee2016accessing, low2009universal, takei2016direct, guardado2018probing, nichols2019spin, lienhard2018observing, hazzard2014many, garttner2017measuring}. The approximations estimate the quantum expectation of observables as the average over classical trajectories of initial phase-space points which are sampled from the Wigner distribution associated with the initial state. They are simple to implement, and offer accuracy consistent with being semiclassical expansions~\cite{schachenmayer2015many, schachenmayer2015dynamics, pucci2016simulation}.

\begin{figure}[t]\centering
\includegraphics[width = 0.7\columnwidth]{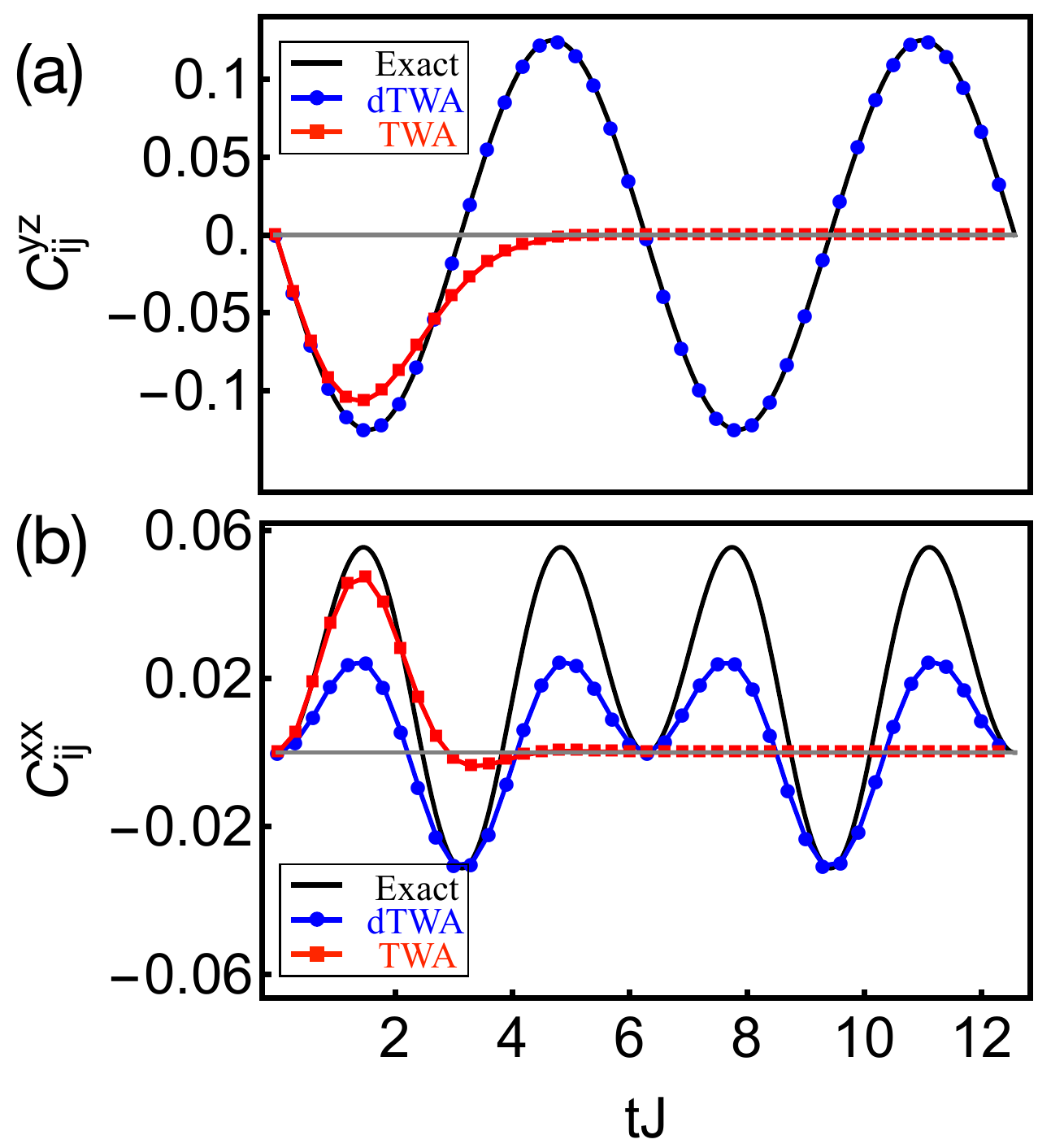}
\caption{(Color online) Dynamics of one component of the spin-spin correlations for the 1D Ising model with no transverse field [whose Hamiltonian is Eq.~\eqref{eqn: HIsing}], obtained from the exact solution (black solid curve), DTWA (blue curve with circles), and TWA (red curve with squares): (a) $C_{ij}^{yz}$ for an initial state with all spins along $\mathbf{x}$, and (b) $C_{ij}^{xx}$ for an initial state with all spins $45^\circ$ between $\mathbf{x}$ and $\mathbf{z}$. $C^{\mu\nu}_{ij}$ is defined in Eq.~\eqref{eqn: C}. The black and blue curves overlap in (a).}
\label{fig: correlation2}
\end{figure}

Earlier works~\cite{schachenmayer2015many, schachenmayer2015dynamics} have argued that DTWA is a superior approximation to calculate the dynamics of spin-spin correlations than TWA, based on specific examples considered. As an example of a case where DTWA is superior, Fig.~\ref{fig: correlation2}(a) shows the dynamics of correlations of neighboring spins in a one-dimensional (1D) Ising chain with no transverse field, obtained from the exact solution, DTWA, and TWA. (The initial conditions and Hamiltonians are described in the figure caption, while the DTWA and TWA calculations will be explained later.) For this case, DTWA exactly captures the dynamics of a specific component of spin correlations, while TWA is accurate for this component only at relatively short times.

However, we must exercise caution when claiming that one method is superior to another based on examples like the ones above, especially because there are nine components, $\langle\hS^{\mu}_i\hS^\nu_j\rangle-\langle\hS^\mu_i\rangle\langle\hS^\nu_j\rangle\ (\mu,\nu\in\{x,y,z\})$, of spin-spin correlations to assess. In contrast to Fig.~\ref{fig: correlation2}(a), Fig.~\ref{fig: correlation2}(b) shows that even for the same model, DTWA performs significantly worse and is qualitatively wrong when we look at a different component of the correlations and a different initial condition (described in the figure caption.) It is often not obvious which correlations, if any, are the most important, especially in dynamics far from equilibrium. Therefore, a more comprehensive comparison of the two Wigner approximations is necessary.

The key finding in this paper is that both DTWA and TWA suppress spin correlations along one direction for a broad class of spin dynamics. We show strong numerical evidence for this, and then rigorously prove this for short times. We also find that the accuracy of DTWA versus TWA is more nuanced than simply one being better than the other. These insights are not readily apparent from looking at plots of the nine Cartesian components of spin-spin correlations. We are able to gain insight into the workings of TWA and DTWA and isolate the nuanced differences between them by utilizing the correlation matrix visualization (CMV) technique, which was recently introduced in Ref.~\cite{mukherjee2018geometric} building on geometrical visualization techniques in Refs.~\cite{kimura2003bloch, byrd2003characterization, jakobczyk2001geometry, tilma2002parametrization, rundle2017simple, bertlmann2008bloch, giraud2015tensor, jevtic2014quantum, dunkl2011numerical, kurzynski2016three, sorensen1984product, halstead1984multipole, donne1997pictorial, philp2005way, merkel2008quantum, dowling1994wigner, harland2012towards, gamel2016entangled, garon2015visualizing, leiner2017wigner}. Correlation matrix visualizations encode all the information contained in spin-spin correlations into three-dimensional shapes and allow us to compare all components of the spin-spin correlations on equal footing.

This article is organized as follows. In Sec.~\ref{sec: Wigner} we introduce TWA and DTWA. In Sec.~\ref{sec: CMV} we describe the tools and metrics that we use to analyze the results of TWA and DTWA. In Sec.~\ref{subsec: Ising} we compare spin-spin correlation dynamics for the exact solution, DTWA, and TWA applied to the Ising model with no transverse field. In Sec.~\ref{subsec: other} we compare spin-spin correlation dynamics calculated with these three methods for the nearest-neighbor 1D transverse Ising and XX models. In Sec.~\ref{sec: rigorous proof} we present a rigorous mathematical argument for one of the key findings in Sec.~\ref{sec: results}, that DTWA and TWA always suppress spin-spin correlations along one direction at short times. We distill the lessons of these comparisons and summarize in Sec.~\ref{sec: conclusions}.

\section{Wigner approximations}\label{sec: Wigner}
\begin{figure}[t]\centering
 \includegraphics[width = 0.7\columnwidth]{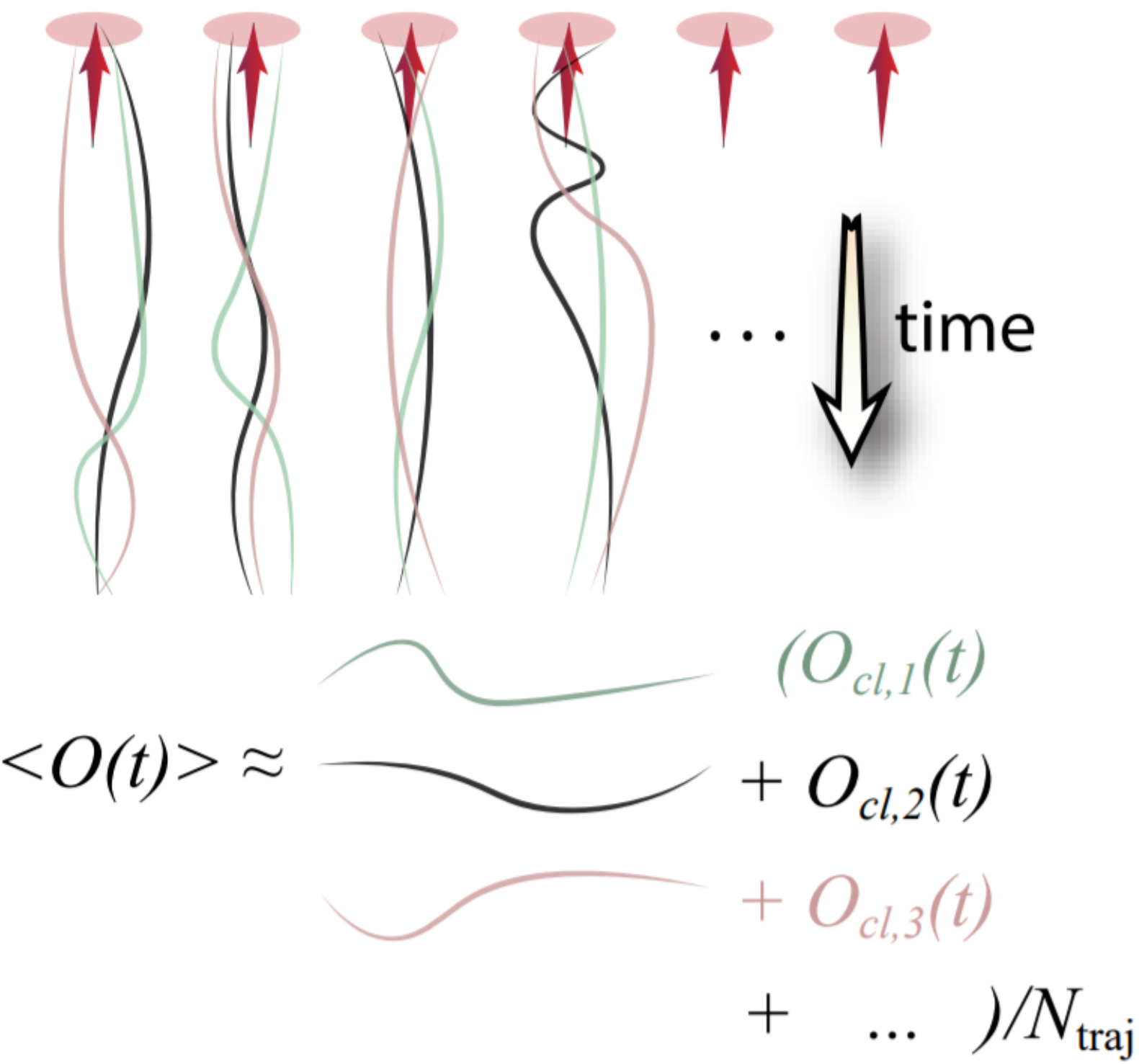}
 \caption{(Color online) Illustration of Wigner approximations. The method consists of three steps: (a) Randomly sample points in phase space from the Wigner distribution for the initial state, (b) evolve the phase points classically through time, and (c) calculate the desired observable from the ensemble average of the observable at time $t$, evaluated from the time-evolved classical trajectories of the initial phase-space points.}
 \label{fig: wigner}
\end{figure}

Wigner approximations approximate dynamics of quantum systems. The implementation of the technique has three steps, schematically illustrated in Fig.~\ref{fig: wigner}. 

In the first step, we sample phase-space coordinates from the Wigner function associated with the initial density matrix $\hat{\rho}(0) = \ket{\psi(0)}\bra{\psi(0)}$. The Wigner function, denoted by $W(\mathbf{S})$, is a quasiprobability distribution that represents $\hat{\rho}(0)$ in an appropriate phase space, with phase points described by coordinates $\mathbf{S}$. The Wigner function $W(\mathbf{S})$ is defined via
\begin{equation}
\hat{\rho} = \int\! d\mathbf{S}~W(\mathbf{S}) \hat{A}(\mathbf{S}), \label{eqn: W}
\end{equation}
where $\hat{A}$ is called a phase-point operator and the integral runs over all of phase space. The phase-space coordinates that describe motional degrees of freedom are position and momentum. For spins, the coordinates can be the spin vector elements $(S^x, S^y, S^z)$. (For spins, the choice of phase space is not unique, and possible phase spaces are discussed in Secs.~\ref{subsec: TWA} and~\ref{subsec: DTWA}.) This step in the algorithm does not contain any approximation, as any observable in a quantum state can be obtained by averaging over phase-space points sampled from the Wigner distribution for that state.

In the second step, we evolve the sampled initial phase-space points in time according to classical equations for the spins. The equations of motion for the specific models we consider [Eqs.~\eqref{eqn: HIsing},~\eqref{eqn: HtransIsing}, and~\eqref{eqn: HXX}] are given in Eqs.~\eqref{eqn: IsingEOM},~\eqref{eqn: transIsingEOM}, and~\eqref{eqn: XXEOM}, respectively. We denote the classical trajectory of an initial point $\mathbf{S}$ by $\mathbf{S}_{\rm cl}(\mathbf{S},t)$.

In the third and final step, we calculate the expectation of an operator $\hat{O}$ at time $t$ by averaging over the trajectories of the phase points as
\begin{equation}\label{eqn: O}
\smallexpect{\hat{O}} = \int\! d\mathbf{S}~{\rm wl}(\hat{O},\mathbf{S}_{\rm cl}(\mathbf{S},t)) W(\mathbf{S}).
\end{equation}
Here, ${\rm wl}(\hat{O},\mathbf{S})$ is the Weyl symbol for $\hat{O}$ at the phase point $\mathbf{S}$. As examples, ${\rm wl}(\hS^\mu_i,\mathbf{S}) = S^\mu_i$ and ${\rm wl}(\hS^\mu_i\hS^\nu_j+\hS^\nu_j\hS^\mu_i,\mathbf{S}) = S^\mu_iS^\nu_j+S^\nu_iS^\mu_j$. The procedure to obtain the Weyl symbol for other observables is more involved~\cite{polkovnikov2010phase}, but in this paper, we only need the examples listed here.

The essence of the Wigner approximations lies in the third step, where we estimate an observable at time $t$ from the classically evolved trajectories of the initial phase-space points. While this step might be intuitive, nevertheless the phase points at time $t$, which are evolved from the initial phase points, \textit{do not} sample the Wigner distribution of the quantum state at $t$. It is for this reason that, sometimes, Wigner approximations give results differing from the exact results. The main purpose of this paper is to explore different cases where the Wigner approximations give results differing from the exact results, extract generic trends regarding how they differ, and give a physical insight for these differences. We focus on spin models in this paper.

Different Wigner approximations differ in their choice of phase space. In this article we focus on two kinds of approximations with two different kinds of phase spaces: TWA samples from a finite continuous area of phase space and DTWA samples from a discrete set of phase points. We describe these schemes in Secs.~\ref{subsec: TWA} and~\ref{subsec: DTWA}, respectively.

\subsection{The TWA}\label{subsec: TWA}
In (continuous) TWA~\cite{polkovnikov2010phase}, the initial values of the spins are allowed to take any value in the continuous phase space spanning the points $(s^x,s^y,s^z)^{\otimes N}$, where $N$ is the number of spins. Reference~\cite{polkovnikov2010phase} derives the Wigner function for the state with all the spins pointing along the $\mathbf{z}$ direction to be
\begin{equation}
W(\mathbf{S}_{\rm tot}) \approx \frac{2}{\pi N}{\rm exp}\left(-\frac{(S^x_{\rm tot})^2+(S^y_{\rm tot})^2}{N/2}\right)\delta\left(S^z_{\rm tot}-N/2\right), \label{eqn: W_TWATot}
\end{equation}
where $S^\mu_{\rm tot} = \sum_i S^\mu_i$. Equation~\eqref{eqn: W_TWATot} is exact in the limit $N\to\infty$. Then the Wigner function for a single spin can be taken to be
\begin{equation}
W(\mathbf{S}_i) = \frac{2}{\pi}e^{-2(S^x_i)^2-2(S^y_i)^2}\delta\left(S^z_i-1/2\right). \label{eqn: W_TWA}
\end{equation}
This is one choice for the single-spin Wigner function that is consistent with Eq.~\eqref{eqn: W_TWATot}; other choices may be possible too. When the system has spins all uniformly pointing along a direction besides $\mathbf{z}$ at the initial time, we first initialize the spins along $\mathbf{z}$ by sampling from Eq.~\eqref{eqn: W_TWA} and then rotate all the spins. We always assume that the spins initially point in the $x$-$z$ plane.

\subsection{The DTWA}\label{subsec: DTWA}
In DTWA~\cite{schachenmayer2015many, schachenmayer2015dynamics}, the initial phase space is chosen to be a discrete set of points $\vec{\alpha}=(\vec{\alpha}_1,\vec{\alpha}_2,..\vec{\alpha}_N)$, where $\vec{\alpha}_i$ is the three-component spin vector for the $i$th spin. As a result, the continuous integral in Eq.~\eqref{eqn: O} is replaced by the sum
\begin{equation}
\expect{O}(t) = \sum_{\vec \alpha}~{\rm wl}(\hat{O},\vec{\alpha}_{\rm cl}(\vec\alpha,t))W_{\vec \alpha},
\end{equation}
where $\vec{\alpha}_{\rm cl}(\vec\alpha,t)$ is the classical trajectory of the initial phase point $\vec\alpha$.

The discrete locations where the initial points $\vec{\alpha}_i$ can lie are nonunique, and different works in the literature have made different choices. For example, Ref.~\cite{schachenmayer2015many} describes the case where the phase space for each spin consists of eight points given by
\begin{align}\label{eqn: S8_alpha}
&\mathbf{S}_1 = \frac{1}{2}(1,1,1),\nonumber\\
&\mathbf{S}_2 = \frac{1}{2}(-1,-1,1),\nonumber\\
&\mathbf{S}_3 = \frac{1}{2}(1,-1,-1),\nonumber\\
&\mathbf{S}_4 = \frac{1}{2}(-1,1,-1),\nonumber\\
&\mathbf{S}_{4+r} = -\mathbf{S}_r\ (1\leq r\leq4).
\end{align}
The phase-point operators are defined as $\hat{A}_{\vec{\alpha}_i} = \frac{1}{2}+\vec{\alpha}_i\cdot\hat{\vec{\sigma}}$, where $\hat{\vec{\sigma}}=(\hat{\sigma}^x,\hat{\sigma}^y,\hat{\sigma}^z)$ is the vector of Pauli matrices $\hat{\sigma}^\mu$ ($\mu=x,y,z$). The phase-point operator for $N$ spins is the product $\hat{A}_{\vec\alpha} = \Pi_i \hat{A}_{\vec{\alpha}_i}$. The Wigner function at $\vec{\alpha}$ is $W_{\vec \alpha} = \frac{1}{2^N}{\rm Tr}(\hat{\rho}\hat{A}_{\vec{\alpha}})$. We initialize the spins by sampling them from the probability distribution $|W_{\vec \alpha}|/\sum_\beta |W_{\vec\beta}|$, and when calculating the dynamics of an operator $\hat{O}$, we multiply its Weyl symbol ${\rm wl}(\hat{O},\vec\alpha)$ by the sign of $W_{\vec\alpha}$.

There is flexibility to choose other discrete sets of points in DTWA. Some of these choices are described in Ref.~\cite{pucci2016simulation}. The dynamics of spin systems sampled from different discrete phase spaces differ, as explored in detail in Ref.~\cite{pucci2016simulation}. While the phase spaces chosen in Ref.~\cite{pucci2016simulation} and other references work well for the models and initial conditions studied there, we find that those phase spaces yield significantly worse results for some of the models and conditions we consider in this paper. Therefore, we use only the phase space comprised of the phase points defined in Eq.~\eqref{eqn: S8_alpha}. For this phase space, the correlations in DTWA are accurate to linear order $O(t)$, although as we explain later, differences from the exact dynamics appear at longer times. We have not explored the question of finding the optimal phase space that will most accurately approximate the dynamics in our study.

\section{Geometric analysis of the spin correlations}\label{sec: CMV}
The connected correlations between a pair of spins $i$ and $j$ are
\begin{equation}\label{eqn: c}
c_{ij}^{\mu\nu} = \expect{\hS_i^\mu \hS_j^\nu}-\expect{\hS_i^\mu}\expect{\hS_j^\nu},\ \mu,\nu\in\{x,y,z,+,-\},
\end{equation}
and their symmetric part is given by
\begin{equation}
C_{ij}^{\mu\nu} = \frac{c_{ij}^{\mu\nu} + c_{ij}^{\nu\mu}}{2},\label{eqn: C}
\end{equation}
where $\hS_j^\pm=\frac{\hS_j^x\pm i\hS_j^y}{2}$. The correlation matrix $C_{ij}$ is a $3\times3$ matrix with components $C^{\mu\nu}_{ij},\ \mu,\nu\in\{x,y,z\}$.

Reference~\cite{mukherjee2018geometric} introduced a geometric tool to visualize $C_{ij}$ using a three-dimensional contour called a CMV. We use this tool to analyze the results of the Wigner approximations. We define the CMV below, and refer the reader to Ref.~\cite{mukherjee2018geometric} for a detailed understanding of the CMV.

We define a function proportional to a homogeneous quadratic polynomial,
\begin{equation}
Q_{ij}(\mathbf{r}) = \frac{\mathbf{r}^T\cdot C_{ij}\cdot\mathbf{r}}{(1+r^2)^{3/2}},
\end{equation}
where $\mathbf{r}$ is a three-dimensional vector. The CMV is the locus of points $\mathbf{r}$ where $Q_{ij}(\mathbf{r})$ has a constant magnitude, $Q_{ij}(\mathbf{r}) = \pm P$. Each sign is assigned a different color. We shade points where $Q_{ij}(\mathbf{r})>0$ as red, and points where $Q_{ij}(\mathbf{r})<0$ as blue. Defining the correlation along the direction $\mathbf{n}$ as $C_{ij}^{nn}=\expect{(\hat{\vec{S_i}}\cdot \mathbf{n}) (\hat{\vec{S_j}}\cdot \mathbf{n})}-\langle\hat{\vec{S_i}}\cdot \mathbf{n}\rangle\langle\hat{\vec{S_j}}\cdot \mathbf{n}\rangle$, the points on the CMV along $\mathbf{n}$ can be obtained by solving the equation $|C_{ij}^{nn}/P| = (1+r^2)^{3/2}/r^2$. This equation has exactly two real solutions for $r$ in the limit that $|C_{ij}^{nn}/P|\gg1$, and these solutions are $r\simeq|C_{ij}^{nn}/P|$ and $r\simeq\sqrt{|P/C_{ij}^{nn}|}$. The size of the CMV along this direction is the difference between these solutions, which is roughly $|C_{ij}^{nn}/P|$. Based on this, we can interpret the size of the CMV along $\mathbf{n}$ as being proportional to $C_{ij}^{nn}$ and therefore the lobes of the CMV point along the eigenvectors of the matrix $C_{ij}$.

\begin{figure}[t]\centering
 \includegraphics[width = 0.8\columnwidth]{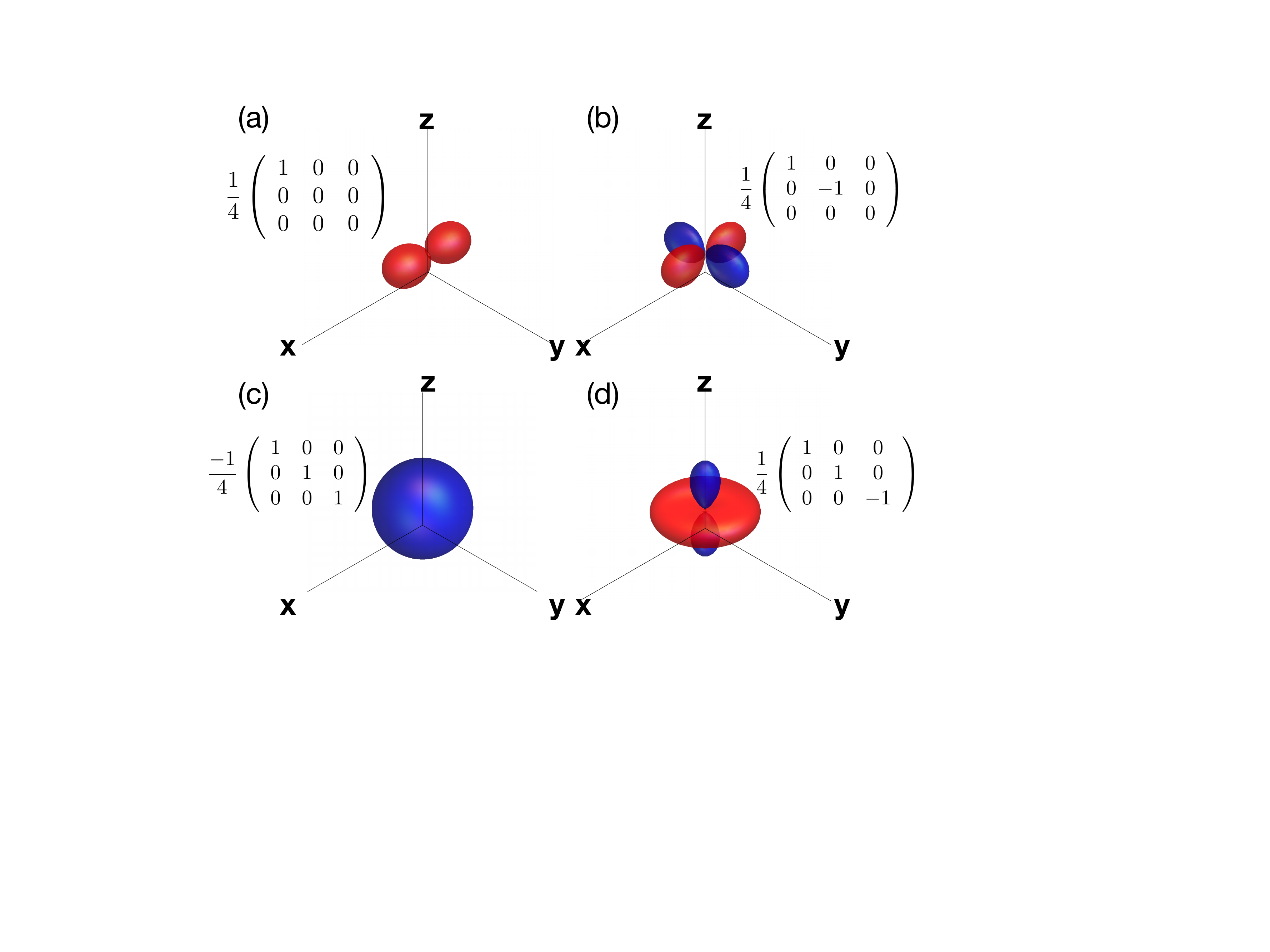}
 \caption{(Color online) Typical CMV shapes for four different cases of the correlation matrix written beside the CMV: (a) A dumbbell, (b) a clover, (c) a sphere, and (d) a wheel and axle.
 }
 \label{fig: CMV}
\end{figure}
We characterize spin-spin correlations via four main features of the CMV. These features are the CMV's size, shape, dimensionality, and orientation. The CMV's size roughly translates to the magnitude of the eigenvalues of $C_{ij}$. The CMV's shape is related to the ratio of the three eigenvalues to each other. The shape generally falls into one of a few categories, depicted in Fig.~\ref{fig: CMV}. When one of the eigenvalues is much larger than the other two, the CMV has the shape of a dumbbell, as in Fig.~\ref{fig: CMV}(a). When two eigenvalues are comparable, have opposite signs, and are larger than the third, the shape is a clover, as in Fig.~\ref{fig: CMV}(b). When all three eigenvalues are comparable, then the shape is a sphere or ellipsoid as in Fig.~\ref{fig: CMV}(c) if they have the same sign, and the shape resembles a wheel and axle as in Fig.~\ref{fig: CMV}(d) if one eigenvalue has a different sign. The CMV's dimensionality is contained in the description of its shape, but this feature is so important in our comparisons that we classify it separately. A dumbbell-shaped CMV is ``one dimensional,'' a clover-shaped one is ``two dimensional,'' and a sphere is ``three dimensional.'' The CMV's orientation tells us the directions of the eigenvectors of $C_{ij}$.

The features described above, despite being qualitative, nevertheless allow us to characterize the differences between Wigner approximations and the exact dynamics, as well as to identify the missing aspects of Wigner approximations. For example, we observe distinct and fairly simple trends such as that DTWA captures the revivals in the size of the CMVs more accurately than TWA (as already shown in Refs.~\cite{schachenmayer2015many, schachenmayer2015dynamics}.) Our most surprising finding is that both DTWA and TWA suppress correlations along one direction, thereby reducing the dimensionality of the CMV. On the other hand, the trends for the accuracy of TWA and DTWA are less apparent in the conventional way of plotting all components of the correlation matrix. Appendix~\ref{sec: component plots} shows the conventional componentwise analysis of correlations for the dynamics considered in the main text, so a curious reader can explore these themselves.

\section{Results}\label{sec: results}
In this section, we compare the dynamics of spin-spin correlations in DTWA, TWA, and the exact solution for various spin models. Specifically, in Sec.~\ref{subsec: Ising} we present the spin dynamics in the nearest-neighbor Ising model with no transverse field, in different dimensions, with different range of interactions, and from different initial states. Section~\ref{subsec: other} presents the spin dynamics in the 1D transverse field nearest-neighbor Ising model and the 1D nearest-neighbor XX model.

\subsection{Ising model}\label{subsec: Ising}
First, we consider the Ising model
\begin{equation}\label{eqn: HIsing}
\hH_I = -\sum_{i\neq j} J_{ij} \hS_i^z\hS_j^z
\end{equation}
with arbitrary interactions $J_{ij}$. 
The time-dependent equations for the quantum-mechanical spin operators are obtained from Heisenberg's equation $i\partial_t\hat{S}^\mu_i = [\hat{S}^\mu_i,\hH]$, resulting in
\begin{align}\label{eqn: IsingEOM}
\dot{\hS}_i^x &= \hS_i^y\hB_i^z,\nonumber\\
\dot{\hS}_i^y &= -\hS_i^x\hB_i^z,\\
\dot{\hS}_i^z &= 0,\nonumber
\end{align}
where $\hB_i^\mu = \sum_{j\neq i} J_{ij}\hS_j^\mu$. The same equations give the classical equations of motion for DTWA and TWA as well, with the quantum-mechanical operator $\hat{S}^\mu_i$ replaced by its classical counterpart $S^\mu_i$.
We initialize the system in the product state $\ket{\theta\theta\theta\textellipsis}$ with $\ket{\theta}=\cos\theta\ket{\uparrow}+\sin\theta\ket{\downarrow}$. We consider two different representative cases in the following sections: $\theta=\frac{\pi}{2}$ and $\theta=\frac{\pi}{4}$. 

First, we will analytically solve this model. Equations~\eqref{eqn: IsingEOM} are integrable, and the solutions are
\begin{equation}\label{eqn: IsingEOMSolns}
\left(\begin{array}{c}\hS^+_j(t)\\ \hS^-_j(t)\\ \hS^z_j(t)\end{array}\right) = \left(\begin{array}{ccc} e^{-i\hB^z_jt}&0&0 \\ 0&e^{i\hB^z_jt}&0 \\ 0&0&1\end{array}\right)
\left(\begin{array}{c}\hS^+_j(0)\\ \hS^-_j(0)\\ \hS^z_j(0)\end{array}\right).
\end{equation}
The time dependence of $\hS^x_j$ and $\hS^y_j$ can be trivially obtained from $\hS^\pm_j$. Note that $\hB^z_j$ commutes with $\hH_I$, and is therefore a constant. Using the relation that $\langle \hS_i^\mu(0)\hS_j^\nu(0)\rangle = \langle \hS_i^\mu(0)\rangle\langle\hS_j^\nu(0)\rangle$ for $i\neq j$ because the spins are initially independent, we obtain the solutions
\begin{align}\label{eqn: explicitSolns}
\langle \hS_j^+(t)\rangle =& \left(\prod_{l\neq j} \langle e^{-iJ_{jl}t\hS_l^z}\rangle\right) \langle \hS_j^+(0)\rangle,\nonumber\\
\langle\hS_j^+(t)\hS_k^+(t)\rangle =& \left(\prod_{l\neq j,k} \langle e^{-i(J_{jl}+J_{kl})t\hS_l^z}\rangle\right) \langle \hS_j^+(0)e^{-iJ_{jk}t\hS_j^z}\rangle \nonumber\\&\times \langle e^{-iJ_{jk}t\hS_k^z}\hS_k^+(0)\rangle, \nonumber\\
\langle\hS_j^+(t)\hS_k^-(t)\rangle =& \left(\prod_{l\neq j,k} \langle e^{-i(J_{jl}-J_{kl})t\hS_l^z}\rangle\right) \langle \hS_j^+(0)e^{iJ_{jk}t\hS_j^z}\rangle \nonumber\\&\times \langle e^{-iJ_{jk}t\hS_k^z}\hS_k^-(0)\rangle, \nonumber\\
\langle\hS_j^+(t)\hS_k^z(t)\rangle =& \left(\prod_{l\neq j,k} \langle e^{-iJ_{jl}t\hS_l^z}\rangle\right) \langle \hS_j^+(0)\rangle \langle e^{-iJ_{jk}t\hS_k^z}\hS_k^z\rangle, \nonumber\\
\langle \hS_j^-\rangle =& \langle\hS_j^+\rangle^*,\nonumber\\
\langle\hS_j^-(t)\hS_k^-(t)\rangle =&\ \langle\hS_j^+(t)\hS_k^+(t)\rangle^*,\nonumber\\
\langle\hS_j^-(t)\hS_k^+(t)\rangle =&\ \langle\hS_j^+(t)\hS_k^-(t)\rangle^*,\nonumber\\
\langle\hS_j^-(t)\hS_k^z(t)\rangle =&\ \langle\hS_j^+(t)\hS_k^z(t)\rangle^*.
\end{align}
The Cartesian components of the magnetization and spin correlations can be obtained from
\begin{align}
&\langle \hS_j^x\rangle = \langle\hS_j^+\rangle + \langle\hS_j^-\rangle,\nonumber\\
&\langle \hS_j^y\rangle = -i(\langle\hS_j^+\rangle - \langle\hS_j^-\rangle),\nonumber\\
&C_{jk}^{xx} = C_{jk}^{++}+C_{jk}^{+-}+C_{jk}^{-+}+C_{jk}^{--},\nonumber\\
&C_{jk}^{xy} = -i(C_{jk}^{++}-C_{jk}^{+-}+C_{jk}^{-+}-C_{jk}^{--}),\nonumber\\
&C_{jk}^{yy} = -(C_{jk}^{++}-C_{jk}^{+-}-C_{jk}^{-+}+C_{jk}^{--}),\nonumber\\
&C_{jk}^{xz} = C_{jk}^{+z}+C_{jk}^{-z},\nonumber\\
&C_{jk}^{yz} = -i(C_{jk}^{+z}-C_{jk}^{-z}),\nonumber\\
&C_{jk}^{\mu\nu} = C_{jk}^{\nu\mu}.
\end{align}

All that remains is to evaluate the expectations in Eq.~\eqref{eqn: explicitSolns} in the exact solution, DTWA, and TWA. In DTWA and TWA, $\langle\hdots\rangle$ should be interpreted as average over the classical phase-space trajectories. Crucially, the explicit results for Eq.~\eqref{eqn: explicitSolns} in DTWA and TWA differ from the exact solution. This is because DTWA and TWA incorrectly estimate averages for products of spin operators on the same site at the initial time. It is worth noting that despite this crucial error, DTWA and TWA still qualitatively capture much of the dynamics of spin correlations, as we will see shortly. The mismatches with the exact solution have simple trends, which we explore in this section. The dynamics in DTWA can be much improved by going to higher order in the BBGKY hierarchy (which also integrates the Heisenberg equations for products of operators $\hS_i^\mu\hS_j^\nu$) and choosing a different phase space (see, e.g, Ref.~\cite{pucci2016simulation}).

We present explicit closed forms of Eq.~\eqref{eqn: explicitSolns} separately for the exact solution, DTWA and TWA in Eqs.~\eqref{eqn: explicitExactSolns},~\eqref{eqn: explicitDTWASolns}, and~\eqref{eqn: explicitTWASolns} in Appendix~\ref{sec: analytical_expns}. Closed forms for the spin correlations in the exact solution have also been calculated in Refs.~\cite{van2013relaxation, hazzard2014quantum}. To numerically evaluate Eqs.~\eqref{eqn: explicitExactSolns},~\eqref{eqn: explicitDTWASolns}, and~\eqref{eqn: explicitTWASolns} for an arbitrary $J_{ij}$ and $\theta$, we assume a chain with $11$ spins and periodic boundaries in the case of 1D models, and a $4\times4$ lattice with periodic boundary conditions for 2D models.

For the other models we consider in Sec.~\ref{subsec: other}, the solutions are more complicated although still integrable~\cite{calabrese2011quantum, calabrese2012quantum1, calabrese2012quantum2}, so we resort to numerically calculating the correlations. We again show that the mismatch between DTWA, TWA, and the exact solution has a simple trend. We also perturbatively calculate $C_{jk}^{\mu\nu}$ at short times in Sec.~\ref{sec: rigorous proof} for arbitrary spin models and rigorously prove our numerical observation.

\subsubsection{Nearest-neighbor 1D Ising model}

\begin{figure}[t]\centering
 \includegraphics[width = \columnwidth]{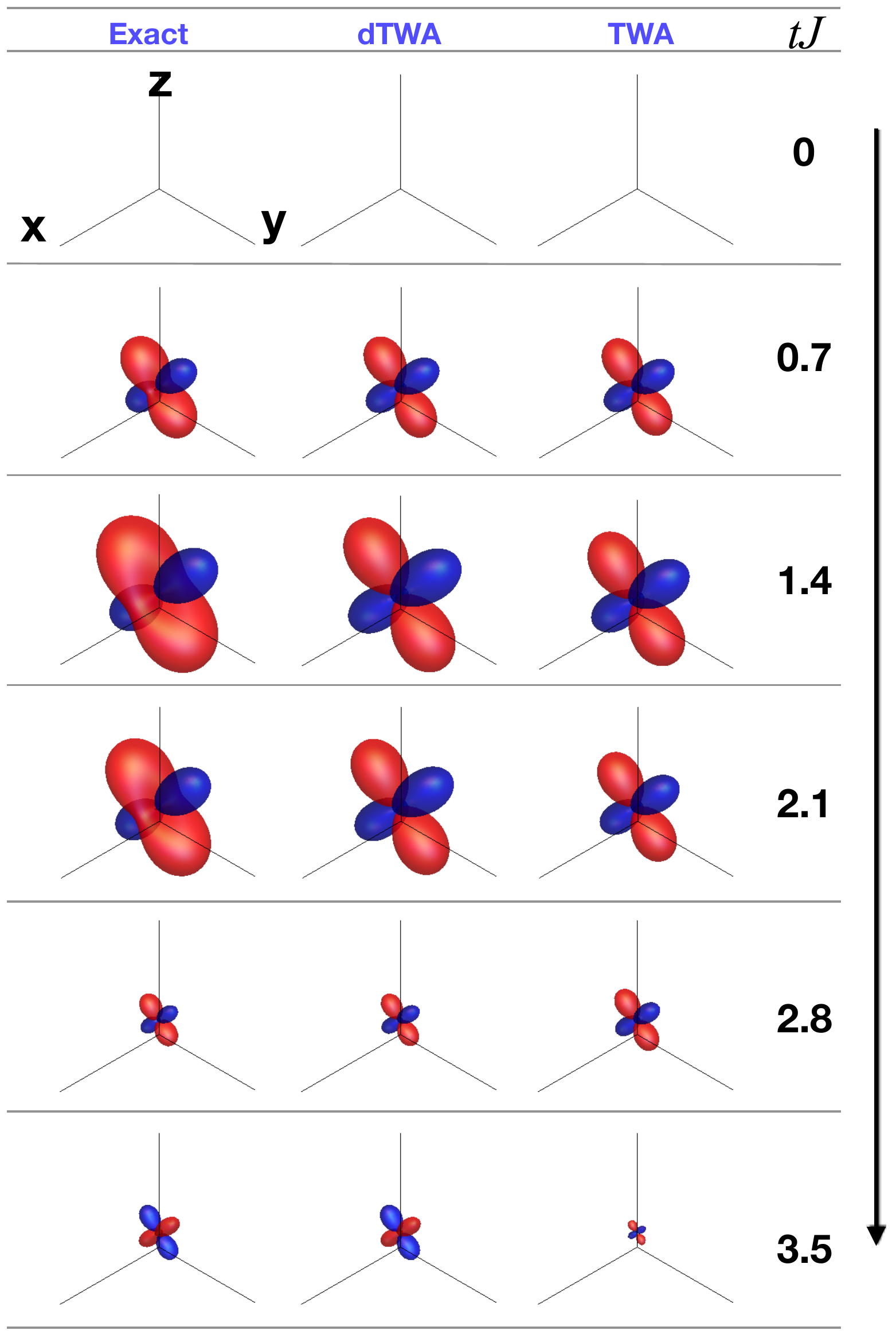}
 \caption{(Color online) The CMVs for nearest-neighbor spin-spin correlations at different times in the nearest-neighbor 1D Ising model in the absence of a transverse field, for the exact solution (left), DTWA (middle), and TWA (right). At $t=0$, all the spins are aligned along $\mathbf{x}$, i.e., $\theta = \frac{\pi}{2}$. An animated movie showing this dynamics is included in the Supplemental Material~\cite{supplement}.}
 \label{fig: Ising}
\end{figure}

First, we study the case $\theta=\pi/2$ and nearest-neighbor interactions in a 1D chain, $J_{ij}=J\delta_{|i-j|=1}$. Figure~\ref{fig: Ising} shows the nearest-neighbor spin correlations for the exact dynamics, DTWA, and TWA.
We find that the shape and orientation of the CMVs are captured well by both TWA and DTWA, and the size is captured well at short times. All the CMVs have a clover shape [as in Fig.~\ref{fig: CMV}(b)]. All the CMVs have the right orientation: They all have large lobes along $\mathbf{y}+\mathbf{z}$ and $\mathbf{y}-\mathbf{z}$.

Despite the similarities listed above, there are two main differences between the exact solution, DTWA, and TWA. The first difference is the well-known inability of TWA to capture the periodic revivals present in the exact solution and DTWA. In fact, DTWA was invented mainly to capture these periodic revivals~\cite{schachenmayer2015many, schachenmayer2015dynamics}. The second difference these results reveal is that in DTWA and TWA, the CMVs are two dimensional, that is, the correlations vanish along the $x$ direction. This can be seen from looking at the components of the correlations in Eq.~\eqref{eqn: explicitTWASolns}. We will see that these differences are general features of spin model dynamics with product state initial conditions.

Our observations in Fig.~\ref{fig: Ising} about the inaccuracies of DTWA and TWA, especially the missing $C^{xx}_{ij}$ correlation, substantiate our argument that it is important to look at all components of the correlations while assessing these approximations. Plotting specific components, as in Fig.~\ref{fig: correlation2}(a), may be misleading about the performance of the approximations. For the model and initial condition considered here, the $C^{yz}$ component in Fig.~\ref{fig: correlation2}(a), which may be viewed as a slice of the CMVs in Fig.~\ref{fig: Ising} along $\frac{\mathbf{y}+\mathbf{z}}{\sqrt{2}}$ (because $C^{yy}_{ij}$ and $C^{zz}_{ij}$ are zero at all times), coincidentally happens to be a component which DTWA captures accurately. These coincidences may not occur for other models or initial conditions, as we will see in the following sections, because the direction misrepresented by the Wigner approximations is often not aligned along a Cartesian direction. All the nonzero Cartesian components of the correlations are plotted in Fig.~\ref{fig: Ising components}.

\begin{figure}[t]\centering
 \includegraphics[width = \columnwidth]{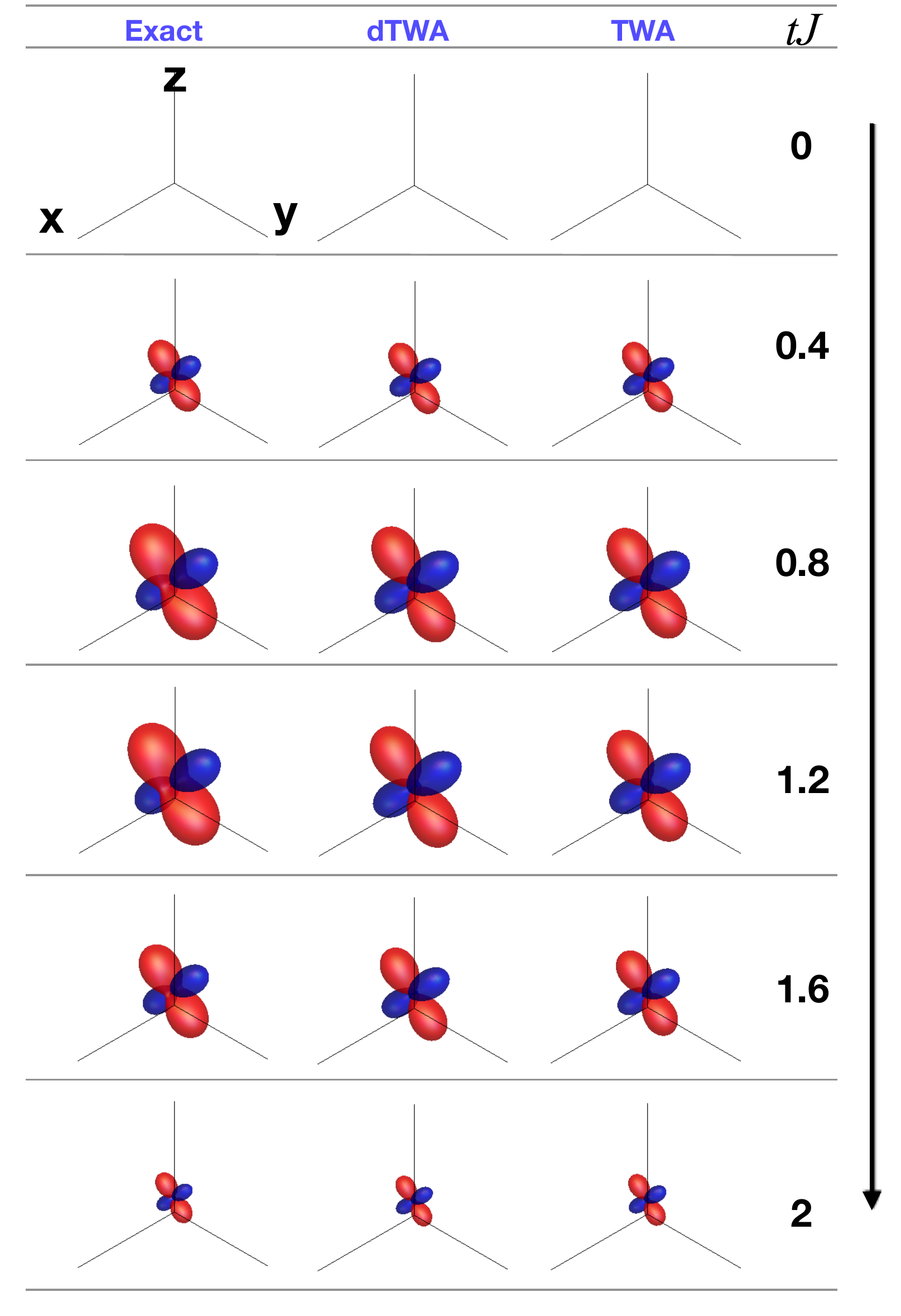}
 \caption{(Color online) The CMVs for nearest-neighbor spin-spin correlations at different times in the nearest-neighbor 2D Ising model in the absence of a transverse field, for the exact solution (left), DTWA (middle), and TWA (right). At $t=0$, all the spins are aligned along $\mathbf{x}$, i.e., $\theta = \frac{\pi}{2}$. An animated movie showing this dynamics is included in the Supplemental Material~\cite{supplement}.}
 \label{fig: 2DIsing}
\end{figure}
\begin{figure}[t]\centering
 \includegraphics[width = \columnwidth]{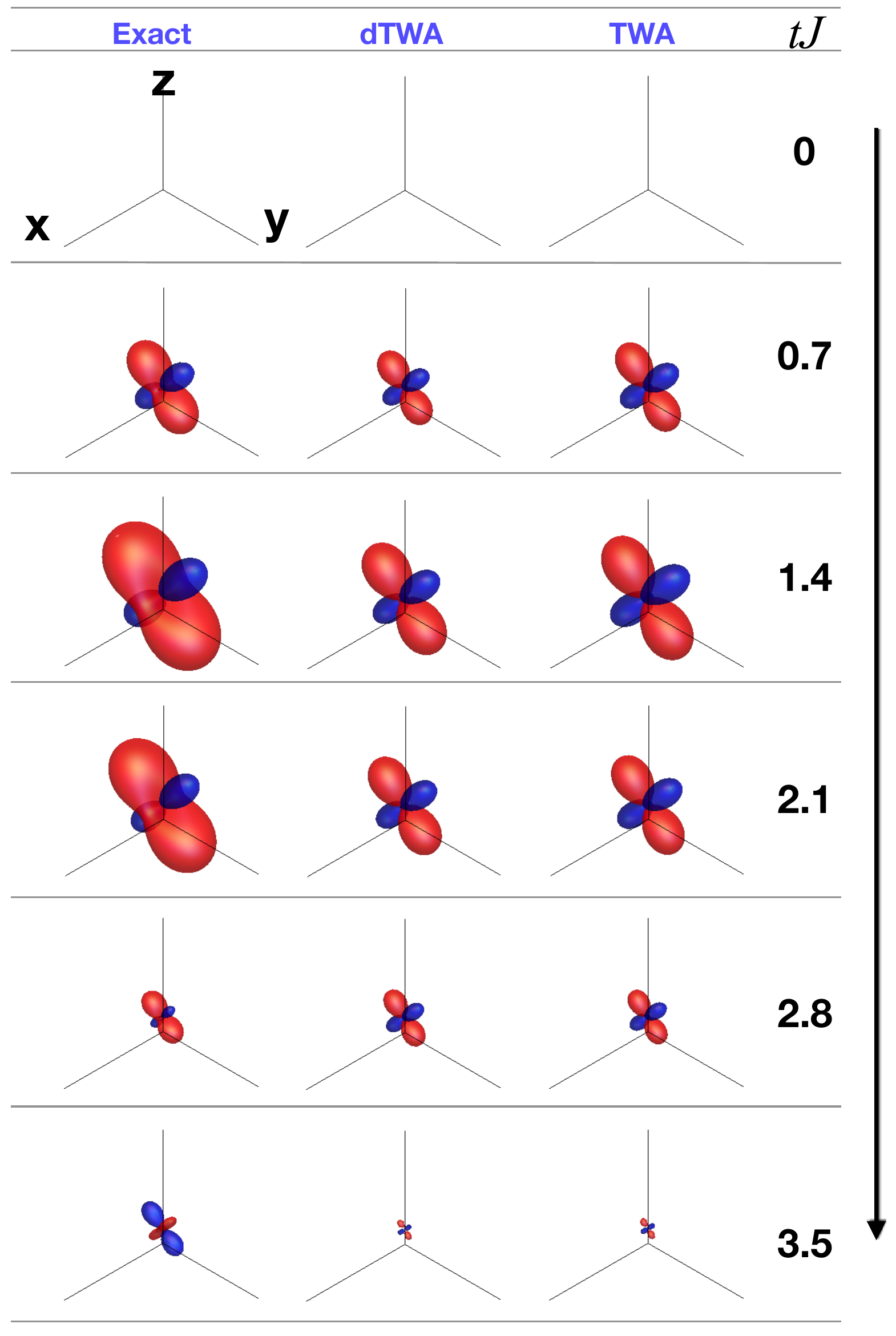}
 \caption{(Color online) The CMVs for nearest-neighbor spin-spin correlations at different times in the 1D Ising model in the absence of a transverse field and $\frac{1}{r^3}$ Ising interaction, for the exact solution (left), DTWA (middle), and TWA (right). At $t=0$, all the spins are aligned along $\mathbf{x}$, i.e., $\theta = \frac{\pi}{2}$. An animated movie showing this dynamics is included in the Supplemental Material~\cite{supplement}.}
 \label{fig: Ising R3}
\end{figure}
\begin{figure}[t]\centering
 \includegraphics[width = \columnwidth]{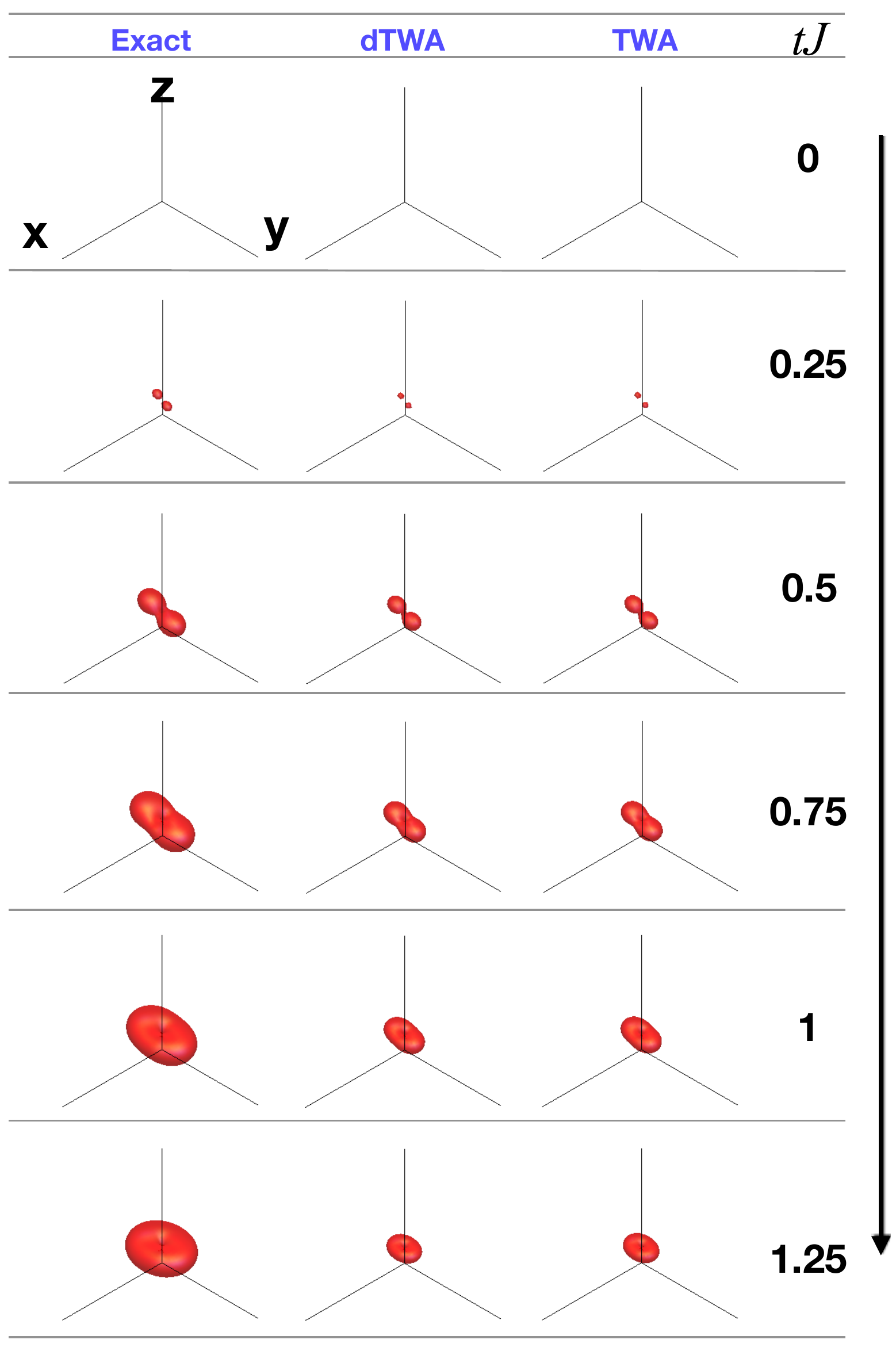}
 \caption{(Color online) The CMVs for spin-spin correlations at different times in the infinite-range Ising model in the absence of a transverse field, for the exact solution (left), DTWA (middle), and TWA (right). At $t=0$, all the spins are aligned along $\mathbf{x}$, i.e., $\theta = \frac{\pi}{2}$. An animated movie showing this dynamics is included in the Supplemental Material~\cite{supplement}.}
 \label{fig: Ising AllInt}
\end{figure}
\begin{figure}[t]\centering
 \includegraphics[width = \columnwidth]{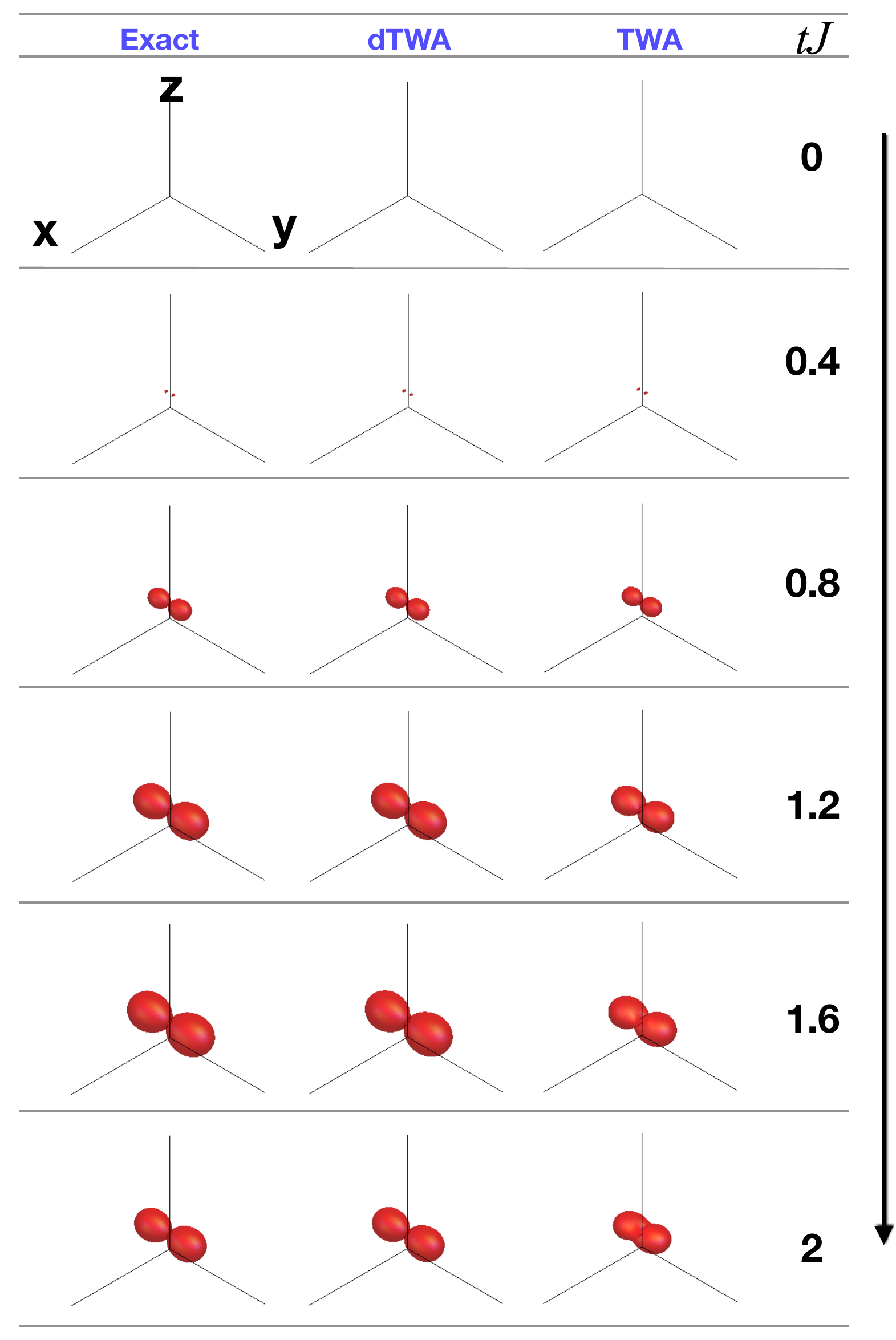}
 \caption{(Color online) The CMVs for next-nearest-neighbor spin-spin correlations at different times in the nearest-neighbor 1D Ising model in the absence of a transverse field, for the exact solution (left), DTWA (middle), and TWA (right). At $t=0$, all the spins are aligned along $\mathbf{x}$, i.e., $\theta = \frac{\pi}{2}$. An animated movie showing this dynamics is included in the Supplemental Material~\cite{supplement}.}
 \label{fig: Ising NNN}
\end{figure}
\begin{figure}[t]\centering
 \includegraphics[width = \columnwidth]{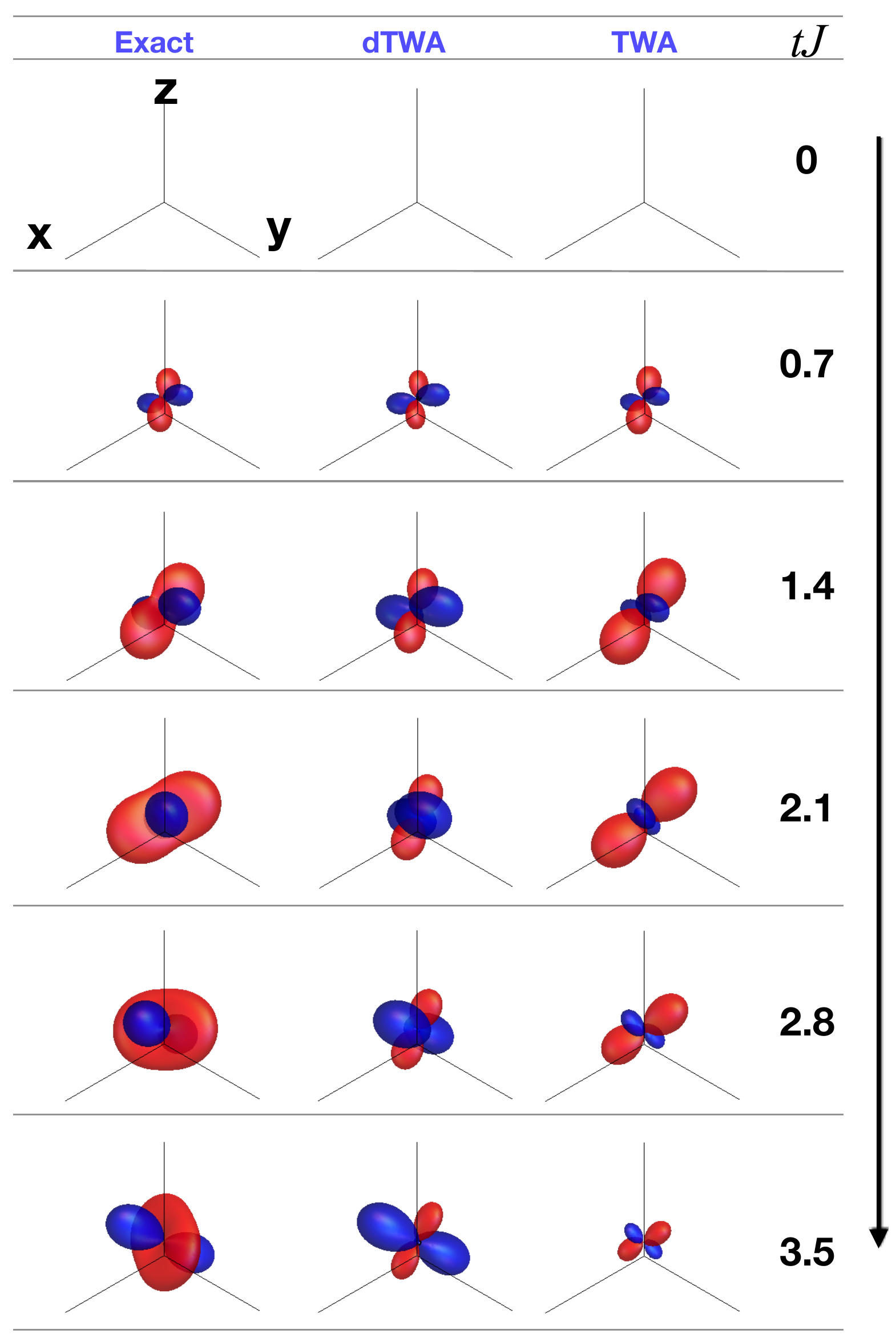}
 \caption{(Color online) The CMVs for nearest-neighbor spin-spin correlations at different times in the nearest-neighbor 1D Ising model in the absence of a transverse field, for the exact solution (left), DTWA (middle), and TWA (right). At $t=0$, all the spins are aligned halfway between $\mathbf{x}$ and $\mathbf{z}$, i.e., $\theta = \frac{\pi}{4}$. An animated movie showing this dynamics is included in the Supplemental Material~\cite{supplement}.}
 \label{fig: Ising pi4}
\end{figure}
\begin{figure}[t]\centering
 \includegraphics[width = \columnwidth]{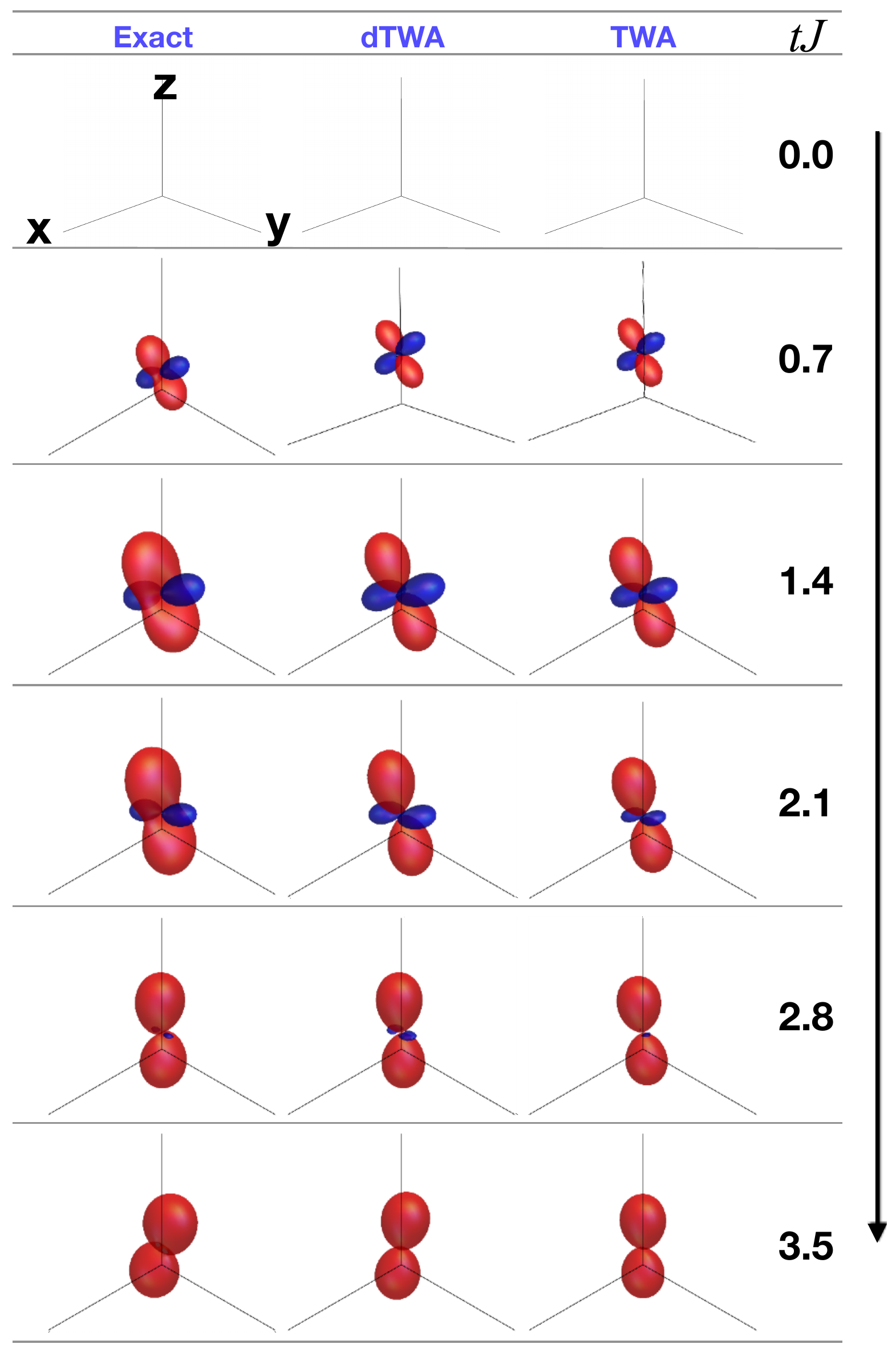}
 \caption{(Color online) The CMVs for nearest-neighbor spin-spin correlations in a nearest-neighbor 1D transverse Ising ($h=J/3$) system at different times, numerically calculated for the exact solution (left), DTWA (middle), and TWA (right). At $t=0$, all the spins are aligned along $\mathbf{x}$, i.e., $\theta = \frac{\pi}{2}$. An animated movie showing this dynamics is included in the Supplemental Material~\cite{supplement}.}
 \label{fig: transIsing}
\end{figure}
\begin{figure}[t]\centering
 \includegraphics[width = \columnwidth]{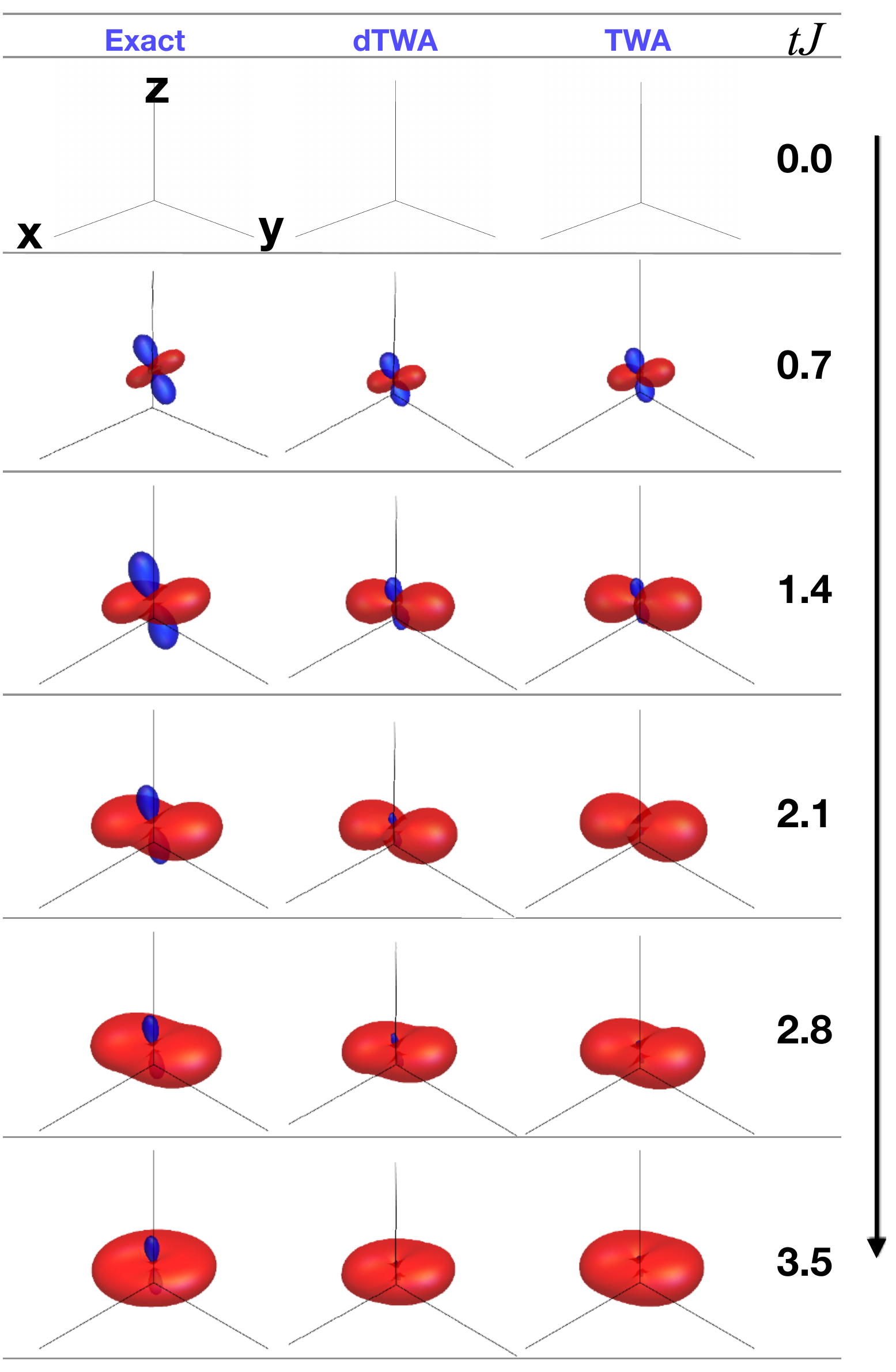}
 \caption{(Color online) The CMVs for nearest-neighbor spin-spin correlations in a 1D system with an $XX$ Hamiltonian, numerically calculated for the exact solution (left), DTWA (middle), and TWA (right), at different times. At $t=0$, all the spins are aligned along $\mathbf{x}$, i.e., $\theta = \frac{\pi}{2}$. An animated movie showing this dynamics is included in the Supplemental Material~\cite{supplement}.}
 \label{fig: XX}
\end{figure}

\subsubsection{Dependence on dimension}\label{subsubsec: 2D}

It is a common expectation that semiclassical approximations perform better in higher dimensions, because the Wigner function does not spread much with time, due to small quantum fluctuations~\cite{polkovnikov2010phase}. To address this, we next study the case $\theta=\pi/2$ and nearest-neighbor interactions in a 2D lattice, $J_{ij}=J\delta_{|\vec{i}-\vec{j}|=1}$.

Figure~\ref{fig: 2DIsing} shows the nearest-neighbor spin correlations for the exact dynamics, DTWA, and TWA. We find that the comparison with the exact solution is similar to the 1D case : The shape and orientation of the CMVs are captured well by both TWA and DTWA, and the size is captured well at short times. Importantly though, the differences in the 1D Ising model also persist in the 2D model: The CMVs in DTWA and TWA are again two dimensional because the correlations completely vanish along $\mathbf{x}$, and the CMVs in TWA exponentially shrink with time. In fact, we rigorously prove in Appendix~\ref{sec: analytical_expns} that the CMV is two dimensional in DTWA and TWA in the nearest-neighbor Ising model in an arbitrary dimension and for any arbitrary initial state. Thus, although going to a higher dimension may improve some aspects of the performance of DTWA or TWA, it does not necessarily remedy the suppression of one correlation component. Further, we show in Sec.~\ref{sec: rigorous proof} that the correlations along the initial Bloch vector in TWA and DTWA are suppressed even for an arbitrary spin model in an arbitrary dimension. All the nonzero Cartesian components of the correlations for this model are plotted in Fig.~\ref{fig: 2DIsing components}.

\subsubsection{Dependence on range of interaction}\label{subsubsec: long-range}

It is also commonly expected that semiclassical approximations perform better for models with long-range interactions, again because the Wigner function does not spread much with time, due to small quantum fluctuations~\cite{polkovnikov2010phase}. To address this, we study two cases: first, Ising interactions decaying as $J_{ij}=\frac{J}{r_{ij}^3}$ in a 1D chain, which is typical in experiments with particles with a dipole moment, and second, infinite-range Ising interactions $J_{ij}=J$, as commonly realized in ion trap experiments. In both cases, we consider the initial state to have $\theta=\pi/2$. The infinite-range Ising model is well studied in the literature and leads to one-axis twisting of the total spin on the Bloch sphere~\cite{kitagawa1993squeezed, ma2011quantum}

Figure~\ref{fig: Ising R3} plots the nearest-neighbor spin correlations for the exact solution, DTWA and TWA in the $1/r^3$ Ising model. These CMVs also have clover shapes [as in Fig.~\ref{fig: CMV}(b)], and are still nearly two dimensional. The component $C^{xx}_{ij}$ is not zero in DTWA and TWA, but is much smaller than it is in the exact solution, as can be observed from the componentwise plots in Fig.~\ref{fig: Ising R3 components}. We will return to a general understanding of this suppression in Sec.~\ref{sec: rigorous proof}. The orientation of the CMVs is captured well by DTWA and TWA, and their size is captured well at short times.

Figure~\ref{fig: Ising AllInt} plots spin-spin correlations for the exact solution, DTWA and TWA in the infinite-range Ising model. Here the DTWA and TWA are capable of reproducing the dynamics at short times. The physical reason for this is that the correlations rapidly develop on a timescale $tJ\sim 1/\sqrt{N}$ (with $N$ being the number of spins), which is faster than the timescale for nearest-neighbor Ising models, essentially because more terms contribute to the dynamics. There is still a small suppression of correlations, but this suppression is much smaller than the magnitude of the correlations, because, as we show in Sec.~\ref{sec: rigorous proof}, the suppression grows on a much slower timescale $tJ\sim 1$. As a result, TWA and DTWA appear to accurately capture the initial rapid growth of correlations. The TWA and DTWA will lead to a noticeable suppression of correlations when $tJ\sim1$, as can be observed in the component wise plots in Fig.~\ref{fig: Ising All components}.

\subsubsection{Dependence on distance between spins}

In the models we study here, correlations in Wigner approximations generally get more accurate as the distance between the two spins increases. Here we calculate the correlations between next-nearest-neighbor spins in the nearest-neighbor 1D Ising model, with spins initialized to $\ket{\theta=\pi/2}$.

Figure~\ref{fig: Ising NNN} shows the next-nearest-neighbor spin correlations for the exact dynamics, DTWA, and TWA. In this case, DTWA agrees perfectly with the exact solution, and this can also be observed in the componentwise plots in Fig.~\ref{fig: Ising NNN components}. The CMVs in the exact solution and DTWA are one dimensional, while the CMVs in TWA are two dimensional, with a small $C^{xx}_{ij}$ component that is absent in the exact solution.

In all nearest-neighbor Ising models in an arbitrary dimension, and with no transverse field as considered throughout this section, all components of the correlations between spins with Manhattan distance greater than $2$ are zero in the exact solution, DTWA, and TWA. This can be easily verified from Eqs.~\eqref{eqn: explicitExactSolns},~\eqref{eqn: explicitDTWASolns}, and~\eqref{eqn: explicitTWASolns}. Correlations between faraway spins are generally not zero in long-range Ising models, and DTWA and TWA are expected to perform well in capturing the dynamics of these long-range correlations as the distance between spins increases. This will get clearer from our rigorous arguments for the dependence of the suppression with distance, which we will present in Sec.~\ref{sec: rigorous proof}.

\subsubsection{Dependence on initial states}

The accuracy and efficiency (i.e., number of samples required) of Wigner approximations depend strongly on the initial state. They become less accurate and significantly more numerically challenging for initial states different from $\theta=\pi/2$ and $\theta=0$. To demonstrate their accuracy, we calculate the nearest-neighbor correlations in the nearest-neighbor 1D Ising model (which is integrable) for $\ket{\theta=\pi/4}$.

Figure~\ref{fig: Ising pi4} shows the nearest-neighbor spin correlations in the exact solution, DTWA, and TWA. The CMVs in both Wigner approximations are again two dimensional at all times, as observed in all nearest-neighbor interaction cases above, and as rigorously proven in Sec.~\ref{sec: rigorous proof} for short times and Appendix~\ref{sec: analytical_expns} for all times. That is, correlations completely vanish along one direction. More interestingly, for this case, the suppressed direction rotates with time (for a closed-form expression of the direction of the vanishing correlations, see Appendix~\ref{sec: analytical_expns}). Aside from the two-dimensionality, the shape of the CMVs in the Wigner approximation reasonably agrees with the exact solution. Again, as expected, the CMVs in TWA exponentially shrinks in size, while the CMVs in DTWA and the exact solution undergo periodic oscillations at a period somewhat longer than the longest time presented in Fig.~\ref{fig: Ising pi4}. Further, there are also hints that the orientation of the CMVs in TWA is closer to the exact solution than the DTWA's is. This is to be expected from looking at Fig.~\ref{fig: correlation2}(b), for example, which showed that even the initial dynamics of $C^{xx}_{ij}$ in DTWA differed significantly from the exact solution and TWA. All the nonzero Cartesian components of the correlations are plotted in Fig.~\ref{fig: Ising pi4 components}.

The real advantage of visualizing the correlations as CMVs is demonstrated by the dynamics considered here: Plotting the CMVs clearly shows that DTWA and TWA completely miss correlations along one eigen direction, a fact which is obscured in the componentwise plots in Fig.~\ref{fig: Ising pi4 components} because the misrepresented direction is not aligned along a Cartesian direction.

For $\theta\notin \{0,\pi/2\}$, we note that DTWA presents a serious numerical obstacle in its implementation: There is a sign problem. The sign problem is notorious in quantum Monte Carlo algorithms, where it arises in fermionic systems as a result of negative wave functions due to anticommutations. The sign problem arises in DTWA because the Wigner function is negative at some of the phase-space points. In these cases, one way to sample the initial points $\mathbf{S}$ in phase space is with the weights $\frac{\left|W(\mathbf{S})\right|}{\int\! d\mathbf{S}~\left|W(\mathbf{S})\right|}$ and then multiply the Weyl symbol for the trajectory of $\mathbf{S}$ by the sign of $W(\mathbf{S})$.

When the sign problem occurs, a sample size scaling exponentially with $N$ is required to obtain a precise ensemble average (i.e with a small sampling error) for any observable in a system with $N$ spins~\footnote{The sampling error for the Bloch vector, i.e., the variance of the sample mean averaged over the classical trajectories, scales as $\alpha^N/N_s$, with $N_s$ the sample size, $N$ the number of spins, and $\alpha = \sum_{S_i}|P(S_i)|$ the sum of absolute values of Wigner functions at the initial phase points for a single spin. When $\theta\neq0,\pi/2$, $\alpha>1$, so the sample error increases exponentially with $N$ for fixed $N_s$.}.
While the results presented in this section were obtained from analytically integrating Eq.~\eqref{eqn: IsingEOM}, which is equivalent to implementing the Wigner approximations with an infinite sample size, a numerical implementation of the Wigner approximations would be computationally expensive. For example, the sampling error for $C^{yy}_{ij}$ at $t=0$ for $\theta=\pi/4$ and a sample size of $10^4$ is $0.019$. This error is comparable to the magnitude of $C^{yy}_{ij}$ during the dynamics and therefore we do not get much useful information about the correlation dynamics. The sampling error for $C^{yy}_{ij}$ reduces to $0.003$ for a larger sample size of $10^5$. This obstacle is not present for $\theta=\pi/2$, where the sampling error for $C^{yy}_{ij}$ for a sample size of $10^4$ is only $0.002$. Other components have similar errors for these sample sizes.
\footnotetext{OK}

The sign problem in DTWA can be ameliorated by rotating the phase space, such that the Wigner function is always positive at the initial phase points that are sampled. However, due to the different alignment between these points and the distinguished directions in the Hamiltonian (e.g., the $\mathbf{z}$ direction in the Ising model), the accuracy of the DTWA would need to be re-evaluated.

\subsubsection{Summary of Ising models}

Based on the integrable examples so far, we are able to observe simple trends regarding Wigner approximations: (a) For nearest-neighbor Ising models on a chain, square, or cubic lattice, the approximations completely miss correlations along one direction relative to the exact solution (this is true on any bipartite lattice, and is rigorously proven in Appendix~\ref{sec: analytical_expns}), (b) for longer-range Ising models, the approximations suppress correlations in the same direction as the nearest-neighbor case at short times, and, as expected, (c) correlations in TWA exponentially decay with time. There are also hints that the correlations are oriented incorrectly in DTWA for initial states different from $\ket{\theta=\pi/2}$. These trends were elegantly captured by plotting CMVs, while they are obscured in the componentwise correlation plots such as Fig.~\ref{fig: correlation2}(b) or Fig.~\ref{fig: Ising pi4 components}. The TWA and DTWA are more accurate in capturing correlations between spins that are far away from each other. The TWA and DTWA also perform better for models with long-range interactions, but their accuracy is limited to shorter times, as can be observed in the infinite-range interaction case. The TWA and DTWA have the same qualitative inaccuracies in nearest-neighbor models in higher dimensions as they do in one dimension.

Next we apply DTWA and TWA to the nearest-neighbor 1D transverse Ising model and the nearest-neighbor 1D XX model. We will find that the discrepancies between the Wigner approximations and the true dynamics have the same qualitative structure as observed in the zero-transverse-field Ising model.

\subsection{XX and transverse Ising models}\label{subsec: other} 

For the nearest-neighbor 1D transverse Ising model given by
\begin{equation}\label{eqn: HtransIsing}
\hat{H}_{\rm T} = \hH_I - h \sum_i \hS^x_i,
\end{equation}
the time-dependent equations for the spins are
\begin{align}\label{eqn: transIsingEOM}
&\dot{\hS}_i^x = \hS_i^y \hB_i^z,\nonumber\\
&\dot{\hS}_i^y = - \hS_i^x \hB_i^z + h \hS_i^z,\\
&\dot{\hS}_i^z = - h \hS_i^y.\nonumber
\end{align}
Equations~\eqref{eqn: transIsingEOM} are not analytically integrable. We numerically integrate them on a periodic chain with 11 spins.

Figure~\ref{fig: transIsing} depicts the CMVs obtained from a numerical implementation of exact diagonalization, DTWA, and TWA, when the system is initialized in $\theta=\pi/2$ and evolves under the model with $h=J/3$. The size, shape, and orientation of the CMVs in TWA and DTWA all approximately match with the exact solution, but as in the $h=0$ cases, the CMVs are somewhat two dimensional in both approximations. That is, the correlation along the direction perpendicular to the obvious clover shape is still much smaller in DTWA and TWA than it is in the exact solution. All the CMVs in these dynamics precess around the magnetic field. All the nonzero Cartesian components of the correlations are plotted in Fig.~\ref{fig: transIsing components}.

For the nearest-neighbor (NN) 1D XX model given by
\begin{equation}\label{eqn: HXX}
\hat{H}_{\rm XX} = - J\sum_{i} (\hS^x_i \hS^x_{i+1} + \hS^y_i \hS^y_{i+1}),
\end{equation}
the time-dependent equations for the spins are
\begin{align}\label{eqn: XXEOM}
&\dot{\hS}_i^x = - \hS_i^z \hB_i^y,\nonumber\\
&\dot{\hS}_i^y = \hS_i^z \hB_i^x,\\
&\dot{\hS}_i^z = \hS_i^x\hB_i^y - \hS_i^y\hB_i^x.\nonumber
\end{align}
Equations.~\eqref{eqn: XXEOM} are not analytically solvable either. We numerically integrate them on a periodic chain with 11 spins.

Figure~\ref{fig: XX} depicts the CMVs obtained from a numerical implementation of exact diagonalization, DTWA, and TWA when the system is initialized in $\theta=\pi/2$. The size, shape, and orientation of the CMVs in TWA and DTWA all approximately match with the exact solution, but the CMVs are again two dimensional in both approximations at short times. Interestingly, at longer times, the direction along which the correlations are dominantly suppressed in DTWA and TWA seems to change somewhat independently of the CMVs' orientations: The CMVs are more squished along $\mathbf{x}$ for $tJ<2.1$ and they are more squished along $\mathbf{z}$ for $tJ>2.1$. All the nonzero Cartesian components of the correlations are plotted in Fig.~\ref{fig: XX components}.

\section{Why do DTWA and TWA suppress correlations?}\label{sec: rigorous proof}

We have observed a suppression of correlations in TWA and DTWA for the Ising, transverse Ising, and XX models. For the $h=0$ Ising models, where we explicitly calculated analytical expressions for the correlations, we attributed the suppression to DTWA and TWA incorrectly estimating averages for initial products of spin operators.
Here we present a general argument that shows that in any spin model for a generic initial product state $\ket{\theta\theta\textellipsis}$, the spin correlation along the initial spin direction $\mathbf{n} = \sin\theta\mathbf{x} + \cos\theta\mathbf{z}$ is always suppressed in DTWA and TWA, at $O(t^2)$. That is, we will show that
\begin{equation}\label{eqn: suppression}
\delta C^{nn}_{ij,\rm DTWA}(t) = |C^{nn}_{ij,\rm exact}(t)| - |C^{nn}_{ij,\rm DTWA}(t)| = At^2 + O(t^3)
\end{equation}
for $A>0$ and similarly for TWA, where $C_{ij}^{nn}$ is the correlation along the initial spin direction, defined as $C_{ij}^{nn} = \mathbf{n}\cdot C_{ij}\cdot\mathbf{n} = \sin^2\theta C_{ij}^{xx} + 2\sin\theta\cos\theta C_{ij}^{xz} + \cos^2\theta C_{ij}^{zz}$. [Note that there is no error to $O(t)$.]

Our argument makes use of the numerical observation that $C_{ij,\rm exact}^{nn}(t)>0$ and $C_{ij,\rm DTWA}^{nn}(t)\geq 0$ at short times. Therefore, to prove Eq.~\eqref{eqn: suppression}, it suffices to show that $C_{ij,\rm exact}^{nn}(t)> C_{ij,\rm DTWA}^{nn}(t)$ at $O(t^2)$.

We consider a general translationally invariant Hamiltonian with two-body interactions,
\begin{equation}
\hat{H} = -\sum_{i\mu} h^\mu \hS_i^\mu - \sum_{i\neq j}\sum_\mu J_{ij}^\mu \hS_i^\mu \hS_j^\mu,
\end{equation}
and the initial product state $\ket{\theta\theta\textellipsis}$ as stated before. This covers all the cases we have considered in this paper.

The time-dependent equation for any spin is
\begin{equation}\label{eqn: arbitrary diffeq}
\dot{\hS}_i^\mu = \epsilon^{\mu\nu\alpha} \hS_i^\nu (h^\alpha + \hB_i^\alpha),
\end{equation}
with $\epsilon$ being the Levi-Civit\`a tensor. We use the Einstein summation convention for the greek indices throughout this section. At short times, $\hS_i^\mu(t)$ is [up to $O(t^2)$]
\begin{align}\label{eqn: Taylor}
\hS_i^\mu(t) = & \hS_i^\mu(0) + t \dot{\hS}_i^\mu + \frac{t^2}{2}\ddot{\hS}_i^\mu \nonumber\\
=& \hS_i^\mu(0) + t \epsilon^{\mu\nu\alpha} \hS_i^\nu(0) [h^\alpha + \hB_i^\alpha(0)]\nonumber\\
 &+ \frac{t^2}{2} \hS_i^\lambda(0) \{ \epsilon^{\nu\lambda\beta}\epsilon^{\mu\nu\alpha} [h^\beta + \hB_i^\beta(0)] [h^\alpha + \hB_i^\alpha(0)]  \nonumber\\
&+  \epsilon^{\mu\lambda\alpha} \epsilon^{\alpha\nu\beta} J_{ij}^\alpha \hS_j^\nu(0) [h^\beta + \hB_j^\beta(0)] \},
\end{align}
where $\ddot{\hS}_i^\mu$ is obtained by differentiating Eq.~\eqref{eqn: arbitrary diffeq}.

We substitute Eq.~\eqref{eqn: Taylor} to calculate $C_{ij}^{\mu\nu}(t)$ in the exact solution, TWA, and DTWA up to $O(t^2)$. We define
\begin{align}
&\mathscr{S}^\mu = \expect{\hS_i^\mu(0)},\nonumber\\
&\mathscr{C}^{\mu\nu}_2 = \expect{\hS_i^\mu(0) \hS_i^\nu(0)},\nonumber\\
&\mathscr{C}^{\mu\nu\lambda}_3 = \expect{\hS_i^\mu(0) \hS_i^\nu(0)\hS_i^\lambda(0)}
\end{align}
 and use the relations
\begin{align}\label{eqn: mathscrC}
& \mathscr{S}^\mu_{\rm exact} = \mathscr{S}^\mu_{\rm DTWA} = \mathscr{S}^\mu_{\rm TWA}, \nonumber\\
& \mathscr{C}^{\mu\nu}_{2,\rm exact} = \frac{1}{4}\delta_{\mu\nu} + \frac{i}{2}\mathscr{S}^\alpha \epsilon^{\mu\nu\alpha}, \nonumber\\
& \mathscr{C}^{\mu\nu}_{2,\rm TWA} = \mathscr{C}^{\mu\nu}_{2,\rm DTWA} = \frac{1}{4}\delta_{\mu\nu}.
\end{align}
We reemphasize that $\mathscr{C}_{2,\rm exact}$ is the quantum expectation of operators, while $\mathscr{C}_{2,\rm DTWA}$ and $\mathscr{C}_{2,\rm TWA}$ are averages over classical trajectories. Note that $\mathscr{C}_3$ can be written similarly to Eq.~\eqref{eqn: mathscrC}, but there are more cases to write, so we do not present them here. 

It is straightforward to show that $C_{ij}^{\mu\nu}(t)$ in the exact solution, TWA, and DTWA are identical to each other at $O(1)$ and $O(t)$. Further, it can be verified, although somewhat tediously, that the difference between the exact solution and the Wigner methods arises at $O(t^2)$, and that the only terms that evaluate to different results are
\begin{align}
C_{ij}^{'\mu\nu}(t) &= t^2 \epsilon^{\mu\mu'\alpha} J_{ij}^\alpha J_{ij}^\beta [ \epsilon^{\nu\nu'\beta}\mathscr{C}^{\mu'\beta}_2\mathscr{C}^{\alpha\nu'}_2 \nonumber\\
& + \frac{1}{2}\epsilon^{\mu'\lambda\beta} \mathscr{S}^\lambda (\mathscr{C}^{\beta\alpha\nu}_3 + \mathscr{C}^{\nu\beta\alpha}_3) \nonumber\\
& +\frac{1}{2}\epsilon^{\alpha\lambda\beta} \mathscr{C}^{\mu'\beta}_2(\mathscr{C}^{\lambda\nu}_2 + \mathscr{C}^{\nu\lambda}_2) ].
\end{align}

The difference between TWA or DTWA and the exact solution can then be evaluated using Eq.~\eqref{eqn: mathscrC}, yielding
\begin{align}
\delta C^{xx}_{ij,\rm DTWA}(t) = &\frac{t^2}{4} \{ (\mathscr{S}^x)^2[(J_{ij}^y)^2+(J_{ij}^z)^2] - (\mathscr{S}^y)^2J^x_{ij}J^y_{ij} \nonumber\\ &- (\mathscr{S}^z)^2J^x_{ij}J^z_{ij} \}, \nonumber\\
\delta C^{xy}_{ij,\rm DTWA}(t) = &\frac{t^2}{4}\{ \mathscr{S}^x\mathscr{S}^yJ_{ij}^z [ J_{ij}^z - 2(\mathscr{S}^z)^2(J_{ij}^x+J_{ij}^y) ] \}.
\end{align}
The other components can be found by cyclic permutation, and $\delta C_{ij,\rm TWA}(t)$ can be similarly obtained from Eq.~\eqref{eqn: mathscrC}.
Specifically, setting $(\mathscr{S}^x,\mathscr{S}^y,\mathscr{S}^z)=\frac{1}{2}(\sin\theta,0,\cos\theta)$,
\begin{align}\label{eqn: error}
\delta C^{nn}_{ij,\rm DTWA}(t) = &\frac{t^2}{16}(J_{ij}^y)^2 + \frac{t^2}{16}( J_{ij}^x\cos^2\theta - J_{ij}^z\sin^2\theta)^2,\nonumber\\
\delta C^{nn}_{ij,\rm TWA}(t) = &\frac{t^2}{16}(J_{ij}^y)^2 + \frac{t^2}{16}( J_{ij}^x\cos^2\theta + J_{ij}^z\sin^2\theta)^2,
\end{align}
which are both nonnegative. This proves that TWA and DTWA always suppress correlations along the initial spin direction at short times, for arbitrary spin models. Our results in this section, which identify the error in TWA and DTWA [Eq.~\eqref{eqn: error}] and their source [Eq.~\eqref{eqn: mathscrC}], could potentially open avenues to modify the semiclassical equations to develop more accurate approximations.

\section{Conclusions}\label{sec: conclusions}
We have demonstrated that the accuracy of Wigner approximations is more nuanced than previously believed, and uncovered properties seemingly intrinsic to both TWA and DTWA, namely, that they incorrectly predict suppressed correlations along one direction. We presented a rigorous perturbative argument to explain the suppressed correlations at short times. The suppressed correlations are often difficult to catch in conventional componentwise plots due to the number and complexity of the correlations and often a misalignment of the suppressed correlation with any Cartesian directions. We also found hints that the orientation of the correlations at short times, at least when the spins do not initially point along a special direction of the Hamiltonian, is sometimes more accurate in TWA than in DTWA. We have systematically explored the performance of DTWA and TWA by changing various parameters, including the dimension of the model, the range of interactions, the distance between the correlated spins, and the initial state, as well as adding external fields to the model, and found that the major source of error in all cases is suppressed correlations along one direction. This observation persists even in cases where semiclassical approximations are expected to work well, such as higher dimensions and long-ranged interactions, as well as other nonintegrable models [such as the 2D transverse Ising model with short- and long-range interactions] that we have studied but not shown in this paper. We have condensed these observations into Table~\ref{table: conclusion}. Understanding the capabilities of TWA and DTWA that we have developed in this paper will better enable practitioners to choose the approximations that are most suited to capture the features they are interested in.

\begin{table}[t]\centering
\begin{tabular}{cccccc}
\hline\hline
{\textbf Model} & Size & Revivals & Shape & 3D nature & Orientation\\
\hline
1D NN Ising & \checkmark & DTWA & \checkmark & $\times$ & \checkmark\\
\hline
2D NN Ising & \checkmark & DTWA & \checkmark & $\times$ & \checkmark\\ 
\hline
1D $\frac{1}{r^3}$ Ising & \checkmark & DTWA & \checkmark & $\times$ & \checkmark\\
\hline
infinite-range Ising & \checkmark & DTWA & \checkmark & $\times$ & \checkmark\\
\hline
NN Ising $\ket{\theta=\frac{\pi}{4}}$ & \checkmark & DTWA & \checkmark & $\times$ & TWA\\
\hline
NN Ising $C_{\langle\langle ij\rangle\rangle}$ & \checkmark & DTWA & \checkmark &  & \checkmark\\
\hline
TIM & \checkmark &  & \checkmark & $\times$ & $\checkmark$\\
\hline
XX & \checkmark &  & \checkmark & $\times$ & \checkmark\\
\hline\hline
\end{tabular}
 \caption{Summary of DTWA's and TWA's abilities in capturing different aspects of spin-spin correlation dynamics in a variety of spin models. We categorize their ability to correctly capture the overall size of CMVs at short times, revival of CMVs at longer times (if applicable), the rough shape up to any suppressed correlations, their 3D nature at short times (i.e., whether DTWA and TWA capture the three-dimensionality of CMVs present in the exact solution), and orientation of CMVs. Any text in the cells means that only the indicated method reasonably captures that category. The DTWA and TWA never have three dimensional CMVs at short times because one correlation component is suppressed in all the cases.}
 \label{table: conclusion}
\end{table}

\section*{Acknowledgments}
This material was based upon work supported with funds from the Welch Foundation, Grant No. C-1872. K.R.A.H. thanks the Aspen Center for Physics, supported by the National Science Foundation Grant No. PHY-1066293, for its hospitality while part of this work was performed. We thank Rick Mukherjee and Anthony Mirasola for useful conversations.

B.S. and K.C.W. contributed equally to this work.

\appendix
\section{Analytical solutions for dynamics in the Ising model}\label{sec: analytical_expns}
Here we use Eq.~\eqref{eqn: explicitSolns} to obtain closed-form solutions for spin correlations in the exact solution, DTWA, and TWA.

\subsection{Exact solution}
To simplify and evaluate Eqs.~\eqref{eqn: explicitSolns} for the exact solution, we use the identity that
\begin{equation} e^{iJ\hS_j^zt}=\cos\frac{Jt}{2}+2i\hS^z_j\sin\frac{Jt}{2}.\end{equation}
Further, for an initial state $\ket{\theta\theta\textellipsis}$, we use the relations $\expect{\vec{\hS}_i} = \frac{1}{2}(\sin\theta,0,\cos\theta)$. Finally, we use the group operations $\hS_j^\mu\hS_j^\nu=i\epsilon_{\mu\nu\lambda}\hS^\lambda_k$. Although familiar, it is important to emphasize these group operations in the exact solution, because they are not true in DTWA and TWA. 

\begin{widetext}
Equations~\eqref{eqn: explicitSolns} yield
\begin{align}\label{eqn: explicitExactSolns}
&\langle \hS_j^+(t)\rangle_{\rm exact} = \frac{1}{4}\sin\theta\prod_{l\neq j} \left(\cos\frac{J_{jl}t}{2}-i\cos\theta\sin\frac{J_{jl}t}{2}\right),\nonumber\\
&\langle \hS_j^z(t)\rangle_{\rm exact} = \frac{1}{2}\cos\theta,\nonumber\\
&\langle \hS_j^+(t)\hS_k^+(t)\rangle_{\rm exact} = \frac{1}{16}\sin^2\theta\prod_{l\neq j,k} \left(\cos\frac{(J_{jl}+J_{kl})t}{2}-i\cos\theta\sin\frac{(J_{jl}+J_{kl})t}{2}\right),\nonumber\\
&\langle \hS_j^+(t)\hS_k^-(t)\rangle_{\rm exact} = \frac{1}{16}\sin^2\theta\prod_{l\neq j,k} \left(\cos\frac{(J_{jl}-J_{kl})t}{2}-i\cos\theta\sin\frac{(J_{jl}-J_{kl})t}{2}\right),\nonumber\\
&\langle \hS_j^+(t)\hS_k^z(t)\rangle_{\rm exact} = \frac{1}{8}\sin\theta \left(\cos\theta\cos\frac{J_{jk}t}{2}-i\sin\frac{J_{jk}t}{2}\right) \prod_{l\neq j,k} \left(\cos\frac{J_{jl}t}{2}-i\cos\theta\sin\frac{J_{jl}t}{2}\right).
\end{align}
\end{widetext}
The special cases given in the text, i.e., the 1D Ising model with nearest-neighbor and long-range interactions, the 2D nearest-neighbor Ising model, and the 1D Ising model with $\theta=\pi/2$ and $\pi/4$, can all be evaluated by a directed substitution of the appropriate $J_{ij}$ and $\theta$. These closed forms were also given in Refs.~\cite{van2013relaxation, hazzard2014quantum}.

\subsection{The DTWA}
In DTWA, the initial spin coordinates are $S_j^\mu=\pm\frac{1}{2}$. Therefore, we again have the identity $e^{iJS_j^zt}=\cos\frac{Jt}{2}+2iS^z_j\sin\frac{Jt}{2}$. However, we do not have the group operations of $(S^x,S^y,S^z)$. In fact, for the choice of phase space in this paper, $\langle S_j^\mu(0)S_j^\nu(0)\rangle = \frac{1}{4}(1-\delta_{\mu\nu})$, where $\delta$ is the Kronecker delta and $\langle\textellipsis\rangle$ refers to the average over the sampled phase points. Using these facts, Eqs.~\eqref{eqn: explicitSolns} yield
\begin{align}
&\langle S_j^+(t)\rangle_{\rm DTWA} = \langle \hS_j^+(t)\rangle_{\rm exact},\nonumber\\
&\langle S_j^z(t)\rangle_{\rm DTWA} = \langle \hS_j^z(t)\rangle_{\rm exact},\nonumber\\
&\langle S_j^+(t)S_k^+(t)\rangle_{\rm DTWA} = \langle \hS_j^+(t)\hS_k^+(t)\rangle_{\rm exact} \cos^2\frac{J_{jk}t}{2}, \nonumber\\
&\langle S_j^+(t)S_k^-(t)\rangle_{\rm DTWA} = \langle \hS_j^+(t)\hS_k^-(t)\rangle_{\rm exact} \cos^2\frac{J_{jk}t}{2}, \nonumber\\
&\langle S_j^+(t)S_k^z(t)\rangle_{\rm DTWA} = \langle \hS_j^+(t)\hS_k^z(t)\rangle_{\rm exact}.
\label{eqn: explicitDTWASolns}
\end{align}
We find that the magnetization in DTWA agrees with the exact solution at all times. However, only the correlation components $C_{jk}^{\mu z}$ and $C_{jk}^{z\mu}$ ($\mu\in\{x,y,z\}$) match with the exact solution, while the components on the $x$-$y$ plane generally agree only at short times.

Two aspects of DTWA are immediately clear from the solutions in Eq.~\eqref{eqn: explicitDTWASolns}. The first is why DTWA performs better in long-range Ising models. The difference between the exact solution [Eq.~\eqref{eqn: explicitExactSolns}] and the DTWA [Eq.~\eqref{eqn: explicitDTWASolns}] is significant only at $t\sim1/J_{jk}$, while the timescale on which correlations initially develop is much faster for long-ranged interactions; for example, for the infinite-range Ising model, correlations develop and the Bloch vector shrinks roughly on a timescale $t\sim1/(J\sqrt{N})$, where $N$ is the total number of spins. Consequently, the discrepancy between the exact solution and DTWA is largest for nearest-neighbor models.

The second aspect that can be observed from Eq.~\eqref{eqn: explicitDTWASolns} is the dimensionality of the CMVs. For example, it can be verified that in a simple toy system with only two spins, the matrix $C_{12}$ always has an eigenvector along the direction of $(\sin\theta,0,\cos\theta\cos\frac{Jt}{2})$ with zero eigenvalue and therefore its CMV is always two dimensional. A similar statement holds true for the nearest-neighbor Ising model in an arbitrary dimension. For the 1D Ising model, the $C_{ij}$ matrix for nearest neighbors $i$ and $j$ has an eigenvector along $\left(1,-\cos\theta\tan\frac{Jt}{2}, \cot\theta(1-\sin^2\theta\sin^2\frac{Jt}{2})\right)$ with a zero eigenvalue and therefore this CMV is two dimensional as well. In the 2D Ising model, the nearest-neighbor $C_{ij}$ matrix has an eigenvector along $\left(\tan\theta(\cos^2\frac{Jt}{2}-3\cos^2\theta\sin^2\frac{Jt}{2})\right.$ ,$-\sin\theta\tan\frac{Jt}{2}(1+2\cos Jt+\sin^2\frac{Jt}{2}\sin^2\theta)$ ,$\left.(1-\sin^2\theta\sin^2\frac{Jt}{2})^3\right)$ with a zero eigenvalue. The next-nearest-neighbor $C_{ij}$ matrix has a zero eigenvalue along $\mathbf{z}$. In contrast, the CMVs for nearest-neighbor correlations in the exact solution are all generally three dimensional.

\subsection{The TWA}
In TWA, the initial phase points for the state $\ket{\theta}$ are obtained by rotating the phase points sampled from the Wigner distribution associated with the state $\ket{\frac{\pi}{2}}$. Thus,
\begin{widetext}
\begin{equation}\label{eqn: twa}
\vec{S}_i(0) = \left(\begin{array}{ccc}\cos\theta&0&\sin\theta\\ 0&1&0\\ -\sin\theta&0&\cos\theta\end{array}\right) \left(\begin{array}{c}X_i/2\\ Y_i/2\\ 1/2\end{array}\right) 
= \frac{1}{2}\left( \sin\theta+X_i\cos\theta, Y_i, \cos\theta-X_i\sin\theta \right)^{\rm T},
\end{equation}
where $X_i$ and $Y_i$ are Gaussian random variables with mean $0$ and variance $1$. Simplifications for $e^{iJS^zt}$ or the group operations of $(S^x,S^y,S^z)$ do not apply here. Therefore, the results in TWA differ from DTWA and the exact solution. Equation~\eqref{eqn: explicitSolns} yields
\begin{align}\label{eqn: explicitTWASolns}
&\langle \hS_j^+(t)\rangle_{\rm TWA} = \frac{1}{4}\sin\theta\prod_{l\neq j} e^{-J_{jl}^2t^2\sin^2\theta/8},\nonumber\\
&\langle \hS_j^z(t)\rangle_{\rm TWA} = \frac{1}{2}\cos\theta,\nonumber\\
&\langle \hS_j^+(t)\hS_k^+(t)\rangle_{\rm TWA} = \frac{1}{16}\sin^2\theta \left(1+i\frac{J_{jk}t\cos\theta}{2}\right)^2 e^{-\frac{J_{jk}^2t^2\sin^2\theta}{4}-iJ_{jk}t\cos\theta} \prod_{l\neq j,k} e^{-\frac{(J_{jl}+J_{kl})^2t^2\sin^2\theta}{8}-i\frac{(J_{jl}+J_{kl})t\cos\theta}{2}},\nonumber\\
&\langle \hS_j^+(t)\hS_k^-(t)\rangle_{\rm TWA} = \frac{1}{16}\sin^2\theta \left(1+\frac{J_{jk}^2t^2\cos^2\theta}{4}\right) e^{-\frac{J_{jk}^2t^2\sin^2\theta}{4}-iJ_{jk}t\cos\theta} \prod_{l\neq j,k} e^{-\frac{(J_{jl}-J_{kl})^2t^2\sin^2\theta}{8}-i\frac{(J_{jl}-J_{kl})t\cos\theta}{2}},\nonumber\\
&\langle \hS_j^+(t)\hS_k^z(t)\rangle_{\rm TWA} = \frac{1}{8}\sin\theta \left(\cos\theta-i\sin^2\theta \frac{J_{jk}t}{2}\right) \prod_{l\neq j,k} e^{-\frac{J_{jl}^2t^2\sin^2\theta}{8}-i\frac{J_{jl}t\cos\theta}{2}}.
\end{align}
\end{widetext}
The magnetization and correlation reasonably (but not exactly) agree with the exact solution at short times and exponentially decay to zero.

Again, two aspects of TWA are immediately clear from Eq.~\eqref{eqn: explicitTWASolns}. The first is that in long-range Ising models and in higher dimensions, the exponential decay $\sim e^{-NJ^2t^2}$ of the correlations in TWA closely mimics the $\sim\cos^N\frac{Jt}{2}$ decay of the correlations in DTWA and exact solution at short times. The second aspect that can be observed is the dimensionality of the CMVs. For a toy system with only two spins, the correlation matrix $C_{12}$ always has an eigenvector along $\left(\cos(\frac{Jt}{2}\cos\theta), -\sin(\frac{Jt}{2}\cos\theta), \cot\theta e^{-J^2t^2\sin^2\theta/8}\right)$ with a zero eigenvalue and therefore its CMV is always two dimensional. For the nearest-neighbor 1D Ising model, the nearest-neighbor $C_{ij}$ matrix always has an eigenvector along $\left(\cos(Jt\cos\theta), -\sin(Jt\cos\theta), \cot\theta e^{-J^2t^2\sin^2\theta/4}\right)$ with zero eigenvalue. In the nearest-neighbor 2D Ising model, the nearest-neighbor $C_{ij}$ matrix always has an eigenvector along $\left(\cos(2Jt\cos\theta), -\sin(2Jt\cos\theta), \cot\theta e^{-J^2t^2\sin^2\theta/2}\right)$ with zero eigenvalue. In contrast, the CMVs for nearest-neighbor correlations in the exact solution are all generally three dimensional.

\begin{figure}[t]\centering
 \includegraphics[width = 1.\columnwidth]{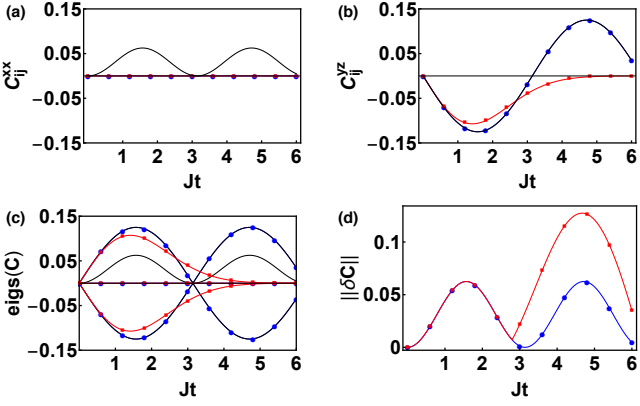}
 \caption{Nearest-neighbor spin correlations for a 1D periodic chain of spins with the nearest-neighbor Ising interaction and initialized to $\ket{\theta=\frac{\pi}{2}}$. (a) and (b) The nonzero components of $C_{ij}$, (c) eigenvalues of $C_{ij}$, and (d) matrix norm of the difference in correlation matrices, $||\delta C_{ij}|| = ||C_{ij,\rm exact} - C_{ij,\rm approx}||$. The black curve shows the exact solution, the blue curve with circles the DTWA, and the red curve with squares the TWA.}
 \label{fig: Ising components}
\end{figure}
\begin{figure}[t]\centering
 \includegraphics[width = 1.\columnwidth]{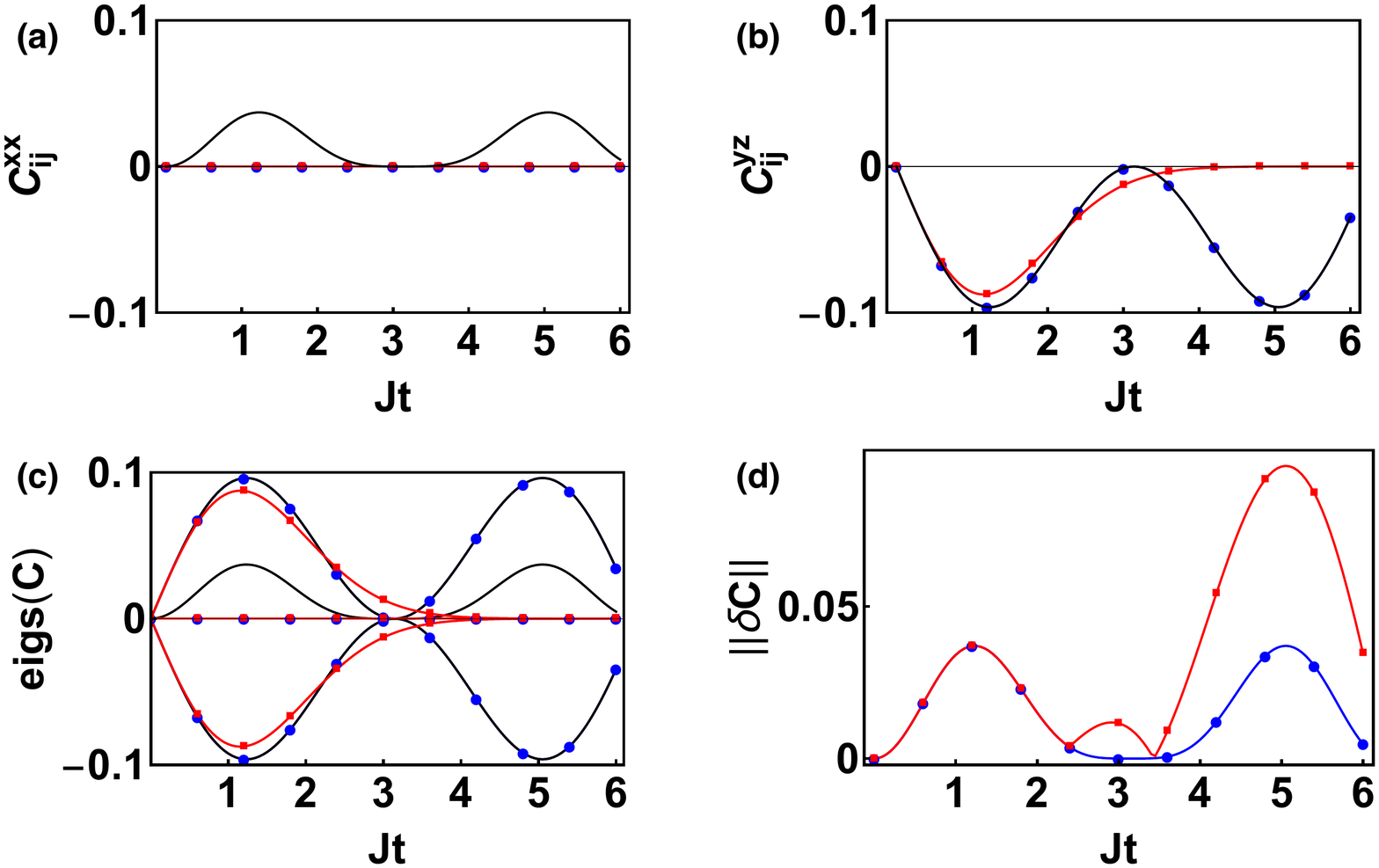}
 \caption{Nearest-neighbor spin correlations for a 2D square lattice of spins with the nearest-neighbor Ising interaction and initialized to $\ket{\theta=\frac{\pi}{2}}$. (a) and (b) The nonzero components of $C_{ij}$, (c) eigenvalues of $C_{ij}$, and (d) matrix norm of the difference in correlation matrices, $||\delta C_{ij}|| = ||C_{ij,\rm exact} - C_{ij,\rm approx}||$. The black curve shows the exact solution, the blue curve with circles the DTWA, and the red curve with squares the TWA.}
 \label{fig: 2DIsing components}
\end{figure}
\begin{figure}[h]\centering
 \includegraphics[width = 1.\columnwidth]{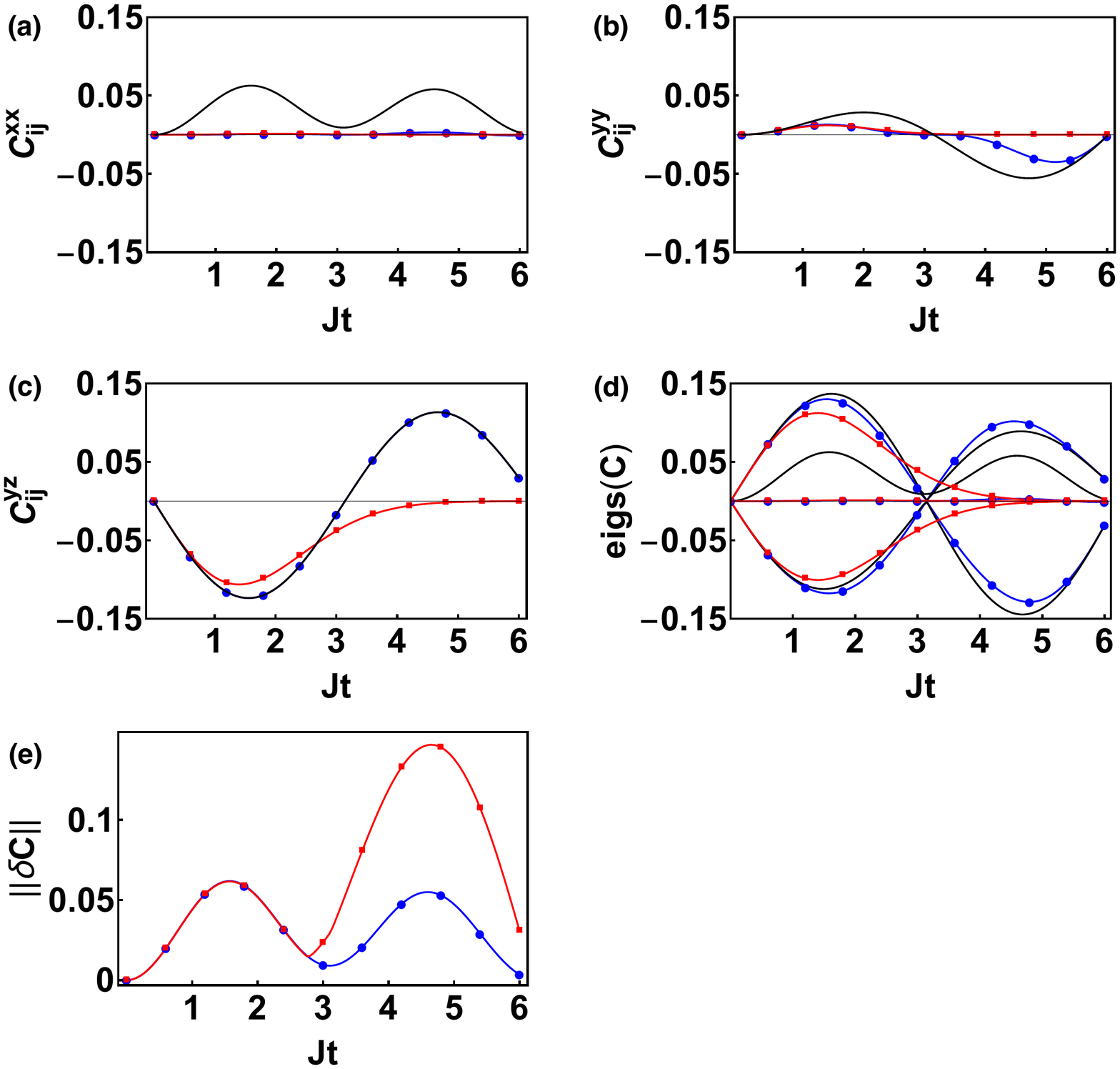}
 \caption{Nearest-neighbor spin correlations for a 1D periodic chain of spins with a long-range Ising interaction decaying with distance as $1/r^3$ and initialized to $\ket{\theta=\frac{\pi}{2}}$. (a)-(c):The nonzero components of $C_{ij}$, (d) eigenvalues of $C_{ij}$, and (e) matrix norm of the difference in correlation matrices, $||\delta C_{ij}|| = ||C_{ij,\rm exact} - C_{ij,\rm approx}||$. The black curve shows the exact solution, the blue curve with circles the DTWA, and the red curve with squares the TWA.}
 \label{fig: Ising R3 components}
\end{figure}
\begin{figure}[h]\centering
 \includegraphics[width = 1.\columnwidth]{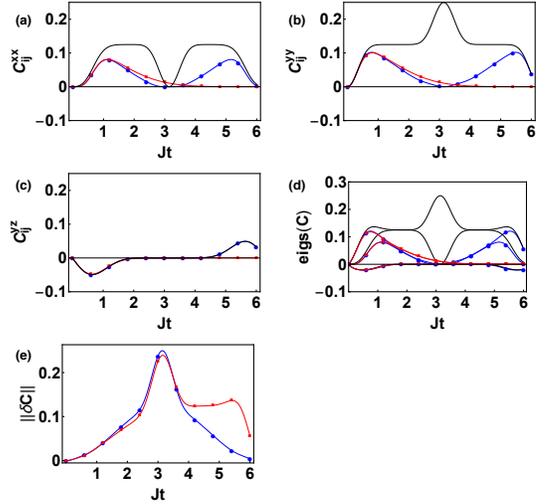}
 \caption{Spin-spin correlations for a systems of spins with infinite-range Ising interaction and initialized to $\ket{\theta=\frac{\pi}{2}}$. (a) and (b) The nonzero components of $C_{ij}$, (c) eigenvalues of $C_{ij}$, and (d) matrix norm of the difference in correlation matrices, $||\delta C_{ij}|| = ||C_{ij,\rm exact} - C_{ij,\rm approx}||$. The black curve shows the exact solution, the blue curve with circles the DTWA, and the red curve with squares the TWA.}
 \label{fig: Ising All components}
\end{figure}
\begin{figure}[h]\centering
 \includegraphics[width = 1.\columnwidth]{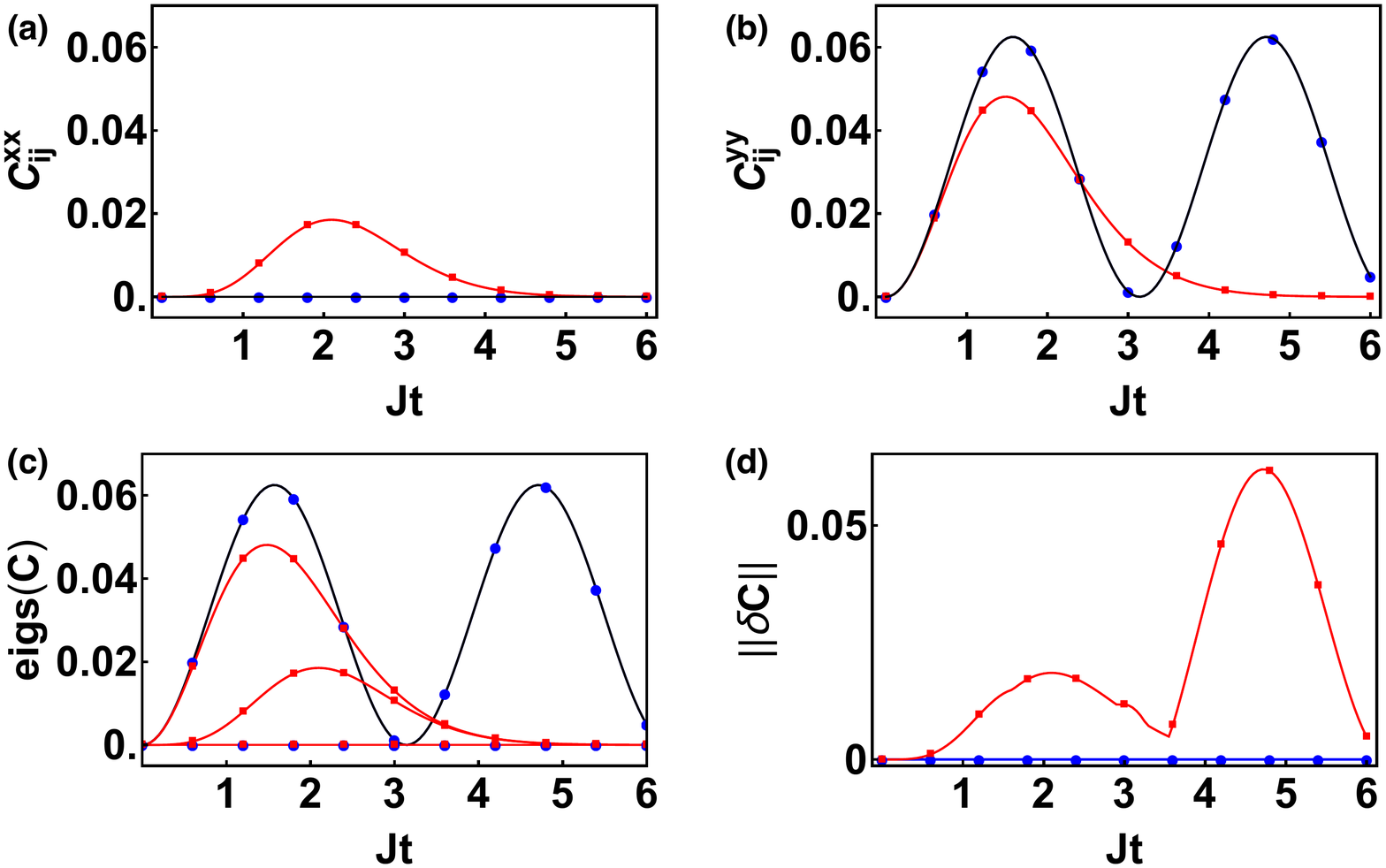}
 \caption{Next-nearest-neighbor spin correlations for a 1D periodic chain of spins with the nearest-neighbor Ising interaction, initialized to $\ket{\theta=\frac{\pi}{2}}$, and interacting with the nearest-neighbor Ising interaction. (a) and (b) The nonzero components of $C_{ij}$, (c) eigenvalues of $C_{ij}$, and (d) matrix norm of the difference in correlation matrices, $||\delta C_{ij}|| = ||C_{ij,\rm exact} - C_{ij,\rm approx}||$. The black curve shows the exact solution, the blue curve with circles the DTWA, and the red curve with squares the TWA.}
 \label{fig: Ising NNN components}
\end{figure}
\begin{figure}[h]\centering
 \includegraphics[width = 1.\columnwidth]{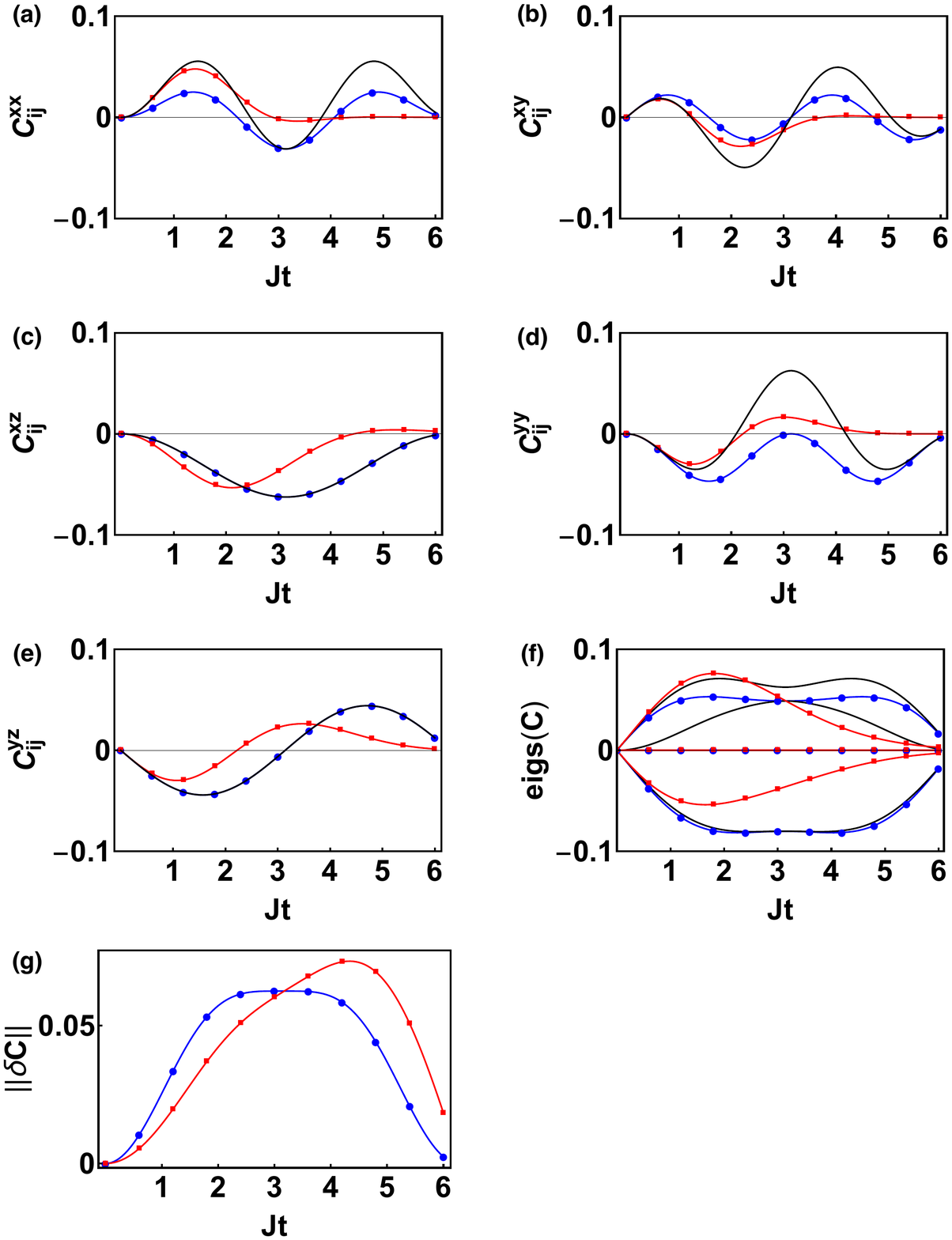}
 \caption{Nearest-neighbor spin correlations for a 1D periodic chain of spins with the nearest-neighbor Ising interaction, initialized to $\ket{\theta=\frac{\pi}{4}}$, and interacting with the nearest-neighbor Ising interaction. (a)-(e) The nonzero components of $C_{ij}$, (f) eigenvalues of $C_{ij}$, and (g) matrix norm of the difference in correlation matrices, $||\delta C_{ij}|| = ||C_{ij,\rm exact} - C_{ij,\rm approx}||$. The black curve shows the exact solution, the blue curve with circles the DTWA, and the red curve with squares the TWA.}
 \label{fig: Ising pi4 components}
\end{figure}
\begin{figure}[h]\centering
 \includegraphics[width = 1.\columnwidth]{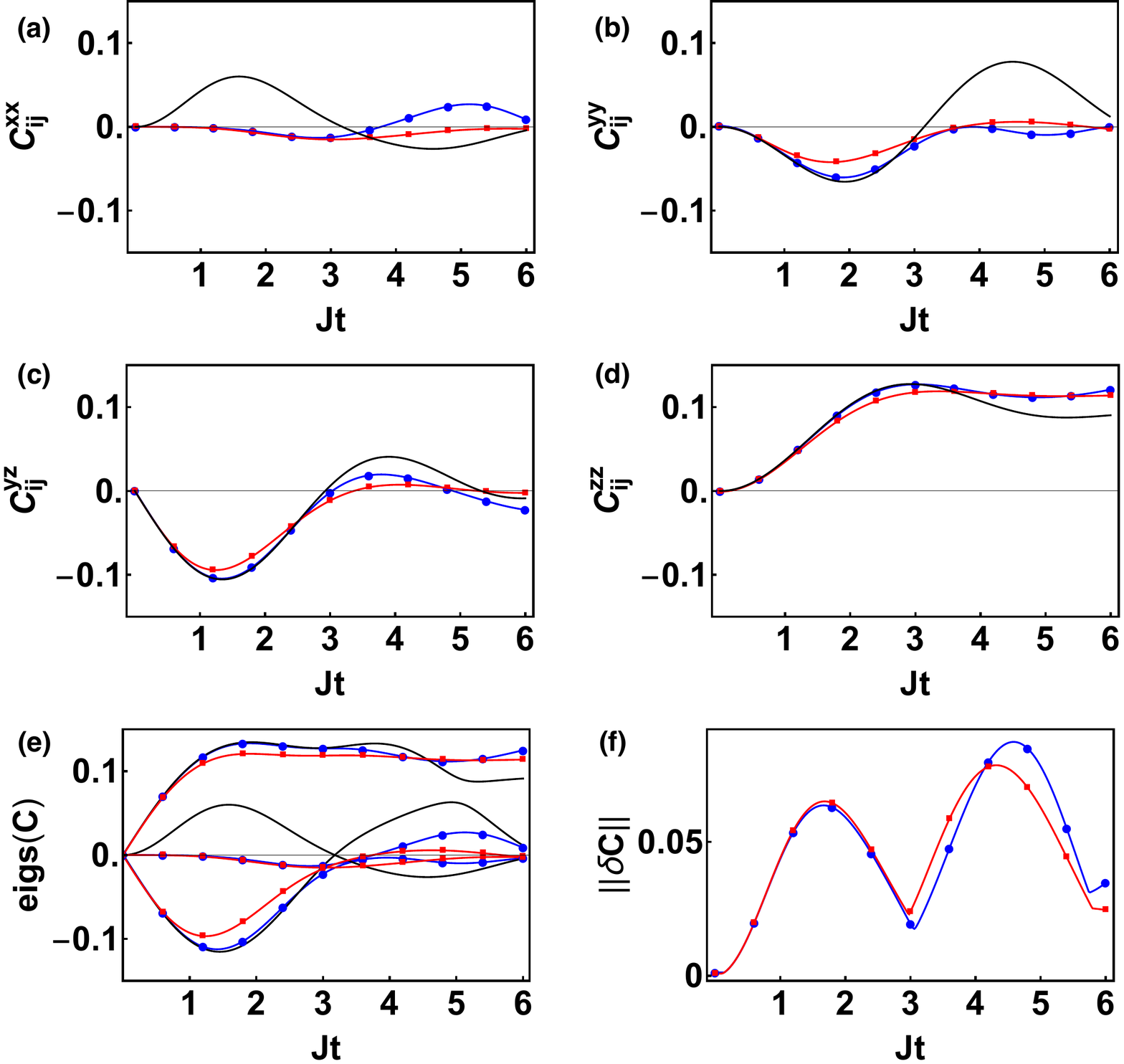}
 \caption{Nearest-neighbor spin correlations for a 1D periodic chain of spins, initialized to $\ket{\theta=\frac{\pi}{2}}$, and interacting with a nearest-neighbor transverse Ising model with $h=J/3$. (a)-(d) The nonzero components of $C_{ij}$, (e) eigenvalues of $C_{ij}$, and (f) matrix norm of the difference in correlation matrices, $||\delta C_{ij}|| = ||C_{ij,\rm exact} - C_{ij,\rm approx}||$. The black curve shows the exact solution, the blue curve with circles the DTWA, and the red curve with squares the TWA.}
 \label{fig: transIsing components}
\end{figure}
\begin{figure}[t]\centering
 \includegraphics[width = 1.\columnwidth]{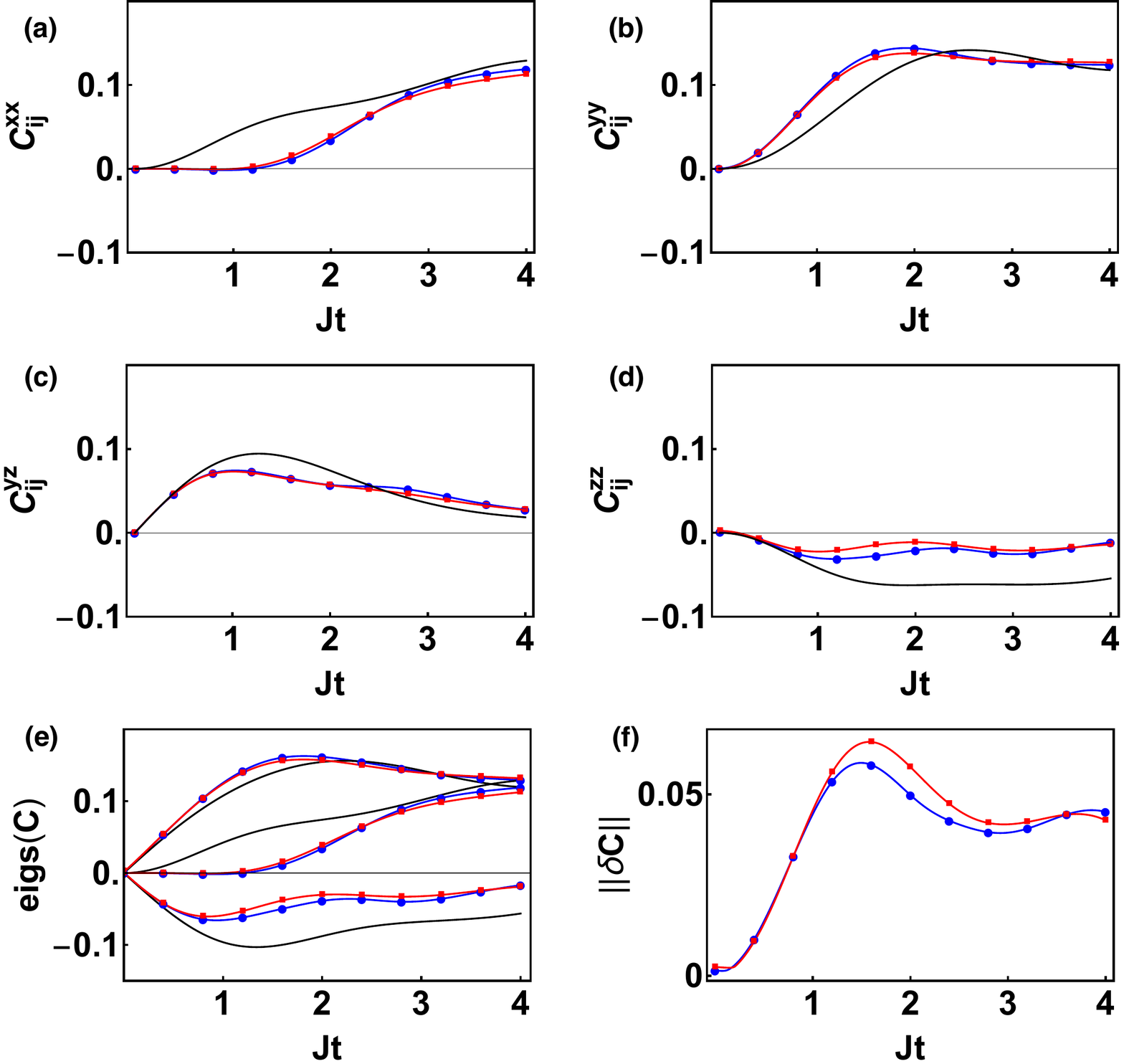}
 \caption{Nearest-neighbor spin correlations for a 1D periodic chain of spins, initialized to $\ket{\theta=\frac{\pi}{2}}$, and interacting with the nearest-neighbor XX model. (a)-(d) The nonzero components of $C_{ij}$, (e) eigenvalues of $C_{ij}$, and (f) matrix norm of the difference in correlation matrices, $||\delta C_{ij}|| = ||C_{ij,\rm exact} - C_{ij,\rm approx}||$. The black curve shows the exact solution, the blue curve with circles the DTWA, and the red curve with squares the TWA.}
 \label{fig: XX components}
\end{figure}

\section{componentwise plots of spin correlations}\label{sec: component plots}
The main text compared TWA and DTWA with the exact solution using CMVs, and several clear observations stood out. For example, the CMVs in the Wigner approximations were two dimensional, vanishing completely in one direction for nearest-neighbor Ising models with no transverse field, and suppressed in all cases (although for infinite-range interactions, the suppression becomes less pronounced as $N\rightarrow \infty$). Moreover, for the initial state $\ket{\theta=\pi/4}$, there were hints that orientation of the CMV was accurate only up to moderate times in DTWA. The CMVs in TWA exponentially shrunk with time for Ising models, as expected.

This appendix presents the same comparisons by conventional means, plotting each Cartesian component separately. Although this is the same information as presented in the main text, it is sometimes less clear from these componentwise plots, or sometimes even completely obscured, what information the Wigner approximations correctly capture or miss, specifically, simple trends such as the correlations along one direction being suppressed in all the Wigner approximations. We also plot the eigenvalues of $C_{ij}$ to directly show that the correlations are suppressed along one direction in DTWA and TWA.

Figures~\ref{fig: Ising components}(a) and~\ref{fig: Ising components}(b) plot all the nonzero Cartesian components of the nearest-neighbor spin correlations for a system initialized in $\ket{\theta=\pi/2}$ and evolving under the 1D Ising model with no transverse field. Figures~\ref{fig: 2DIsing components}(a)-(b) and~\ref{fig: Ising R3 components}(a)-(c) plot the nonzero components for the same initial state, and evolving under the 2D Ising model with no transverse field and the long-range $1/r^3$ 1D Ising model, respectively. For these cases, the figures clearly show that DTWA and TWA suppress the correlations along $\mathbf{x}$, because the suppressed component happens to be along a Cartesian direction. The last two plots in each figure show the eigenvalues of $C_{ij}$ and the matrix norm of the difference in correlation matrices, $||\delta C_{ij}|| = ||C_{\rm exact} - C_{\rm DTWA}||$ and $||\delta C_{ij}|| = ||C_{\rm exact} - C_{\rm TWA}||$, two indicators of the difference between DTWA, TWA, and the exact solution.

Figure~\ref{fig: Ising All components} plots the nonzero Cartesian components of the spin-spin correlations, their eigenvalues, and $||\delta C||$, for the same initial state and evolving under the infinite-range Ising model. The correlations build up rapidly at short times on account of the infinite range of the interaction, and DTWA and TWA agree well with the exact solution at short times. However, DTWA and TWA do not capture any of the dynamics at longer times $tJ\sim\pi$. Figure~\ref{fig: Ising All components}(d) shows that DTWA and TWA obtain the eigenvalues of $C_{ij}$ reasonably well.

Figure~\ref{fig: Ising NNN components} plots the nonzero components of the next-nearest-neighbor correlation, their eigenvalues, and $||\delta C||$, for the same initial state and evolving under the nearest-neighbor 1D Ising model. While there is only nonzero component for the exact solution and DTWA (which captures the exact solution accurately), there are two nonzero components in TWA. The TWA overestimates one of the components, and therefore one of the eigenvalues, of $C_{ij}$.

Figure~\ref{fig: Ising pi4 components} plots the nonzero Cartesian components of the nearest-neighbor spin correlations, their eigenvalues, and $||\delta C||$, for the initial state $\ket{\theta=\pi/4}$ and evolving under the nearest-neighbor 1D Ising model with no transverse field. In contrast to all the cases above, where the suppressed correlations in DTWA and TWA could be clearly observed in the componentwise plots, it is nontrivial in this case to deduce that the correlations are suppressed along one direction from looking at the componentwise plots. The fact that correlations are completely suppressed along one direction is noticeable only by plotting the eigenvalues of $C_{ij}$ in Fig.~\ref{fig: Ising pi4 components}(f), and even this plot is not helpful in arriving at a physical explanation for where and why the correlation is suppressed. On the other hand, the CMVs in Fig.~\ref{fig: Ising pi4} immediately show that DTWA and TWA again completely suppress correlations along one direction, that this direction is aligned with the spins at short times, and that the suppressed direction then precesses with time, all of which is obscured by the componentwise plots.

Figure~\ref{fig: transIsing components} plots all the nonzero Cartesian components of the nearest-neighbor spin correlations, their eigenvalues, and $||\delta C||$, for the initial state $\ket{\theta=\pi/2}$ evolving under the 1D transverse Ising model with $h=J/3$. The plots show that the dominant error in DTWA and TWA is again to partially suppress the correlations along $\mathbf{x}$. The correlations also precess about $\mathbf{x}$, a fact which is not visible from Fig.~\ref{fig: transIsing components} but is immediately apparent in Fig.~\ref{fig: transIsing}. Figure~\ref{fig: XX components} plots all the nonzero Cartesian components of the nearest-neighbor spin correlations, their eigenvalues, and $||\delta C||$, for the initial state $\ket{\theta=\pi/2}$ evolving under the 1D XX model. The correlations in DTWA and TWA are suppressed along $\mathbf{x}$ for $tJ\lesssim2$ and along $\mathbf{z}$ for $tJ\gtrsim2$.

\bibliography{BibFile}
\end{document}